\documentclass[twocolumn]{aa}
\usepackage[colorlinks=true]{hyperref}
\usepackage{graphicx}
\usepackage{txfonts}
\usepackage{tabularx}
\usepackage{amsmath}
\usepackage{xcolor}
\usepackage{cancel}
\usepackage{ulem}
 \usepackage{threeparttable}

\begin{document}

\title {The bright black hole X-ray binary 4U 1543--47 during 2021 outburst: a thick accretion disk inflated by high luminosity}
\titlerunning{4U 1543--47 during 2021 outburst}

\author
{
S. J. Zhao\inst{1,2} 
\and L. Tao\inst{1}\thanks{E-mail: taolian@ihep.ac.cn}
\and P. P. Li\inst{1,2}
\and R. Soria\inst{2,3,4}
\and H. Feng\inst{5}
\and Y. X. Zhang\inst{6,7}
\and R. C. Ma\inst{1,2}
\and W. D. Zhang\inst{8}
\and E. L. Qiao\inst{2,8}
\and Q. Q. Yin\inst{1}
%\and J. L. Qu\inst{1}
\and S. N. Zhang\inst{1,2}
\and L. Zhang\inst{1}
\and Q. C. Bu\inst{9}
\and X. Ma\inst{1}
\and Y. Huang\inst{1}
\and M. Y. Ge\inst{1}
\and X. B. Li\inst{1}
\and L. Chen\inst{10}
\and Q. C. Zhao\inst{1,2}
\and J. Q. Peng\inst{1,2}
\and Y. X. Xiao\inst{1,2}
}
% List of institutions

\institute{Key Laboratory of Particle Astrophysics, Institute of High Energy Physics, Chinese Academy of Sciences, Beijing 100049, China
\and University of Chinese Academy of Sciences, Chinese Academy of Sciences, Beijing 100049, China
\and INAF-Osservatorio Astrofisico di Torino, Strada Osservatorio 20, I-10025 Pino Torinese, Italy
\and Sydney Institute for Astronomy, School of Physics A28, The University of Sydney, Sydney, NSW 2006, Australia
\and Department of Astronomy, Tsinghua University, Beijing 100084, China
\and Kapteyn Astronomical Institute, University of Groningen, P.O.\ BOX 800, 9700 AV Groningen, The Netherlands
\and Center for Astrophysics, Harvard \& Smithsonian, 60 Garden St, Cambridge, MA 02138, USA
\and Key Laboratory of Space Astronomy and Technology, National Astronomical Observatories, Chinese Academy of Sciences, Being
100012, China
\and Institut f{\"u}r Astronomie und Astrophysik, Sand 1, 72076 T{\"u}bingen, Germany
\and Shanghai Astronomical Observatory, Chinese Academy of Sciences, 80 Nandan Road, Shanghai 200030, China
}

\authorrunning{S.J.Zhao et al.}
\date{Received ; accepted }

% \abstract{}{}{}{}{} 
% 5 {} token are mandatory
 
  \abstract 
   {The black hole X-ray binary source 4U 1543--47 experienced a super-Eddington outburst in 2021, reaching a peak flux of up to $\sim1.96\times10^{-7}\rm erg\ \rm cm^{-2}\ \rm s^{-1}$ ($\sim 8.2$ Crab) in the 2--10\,keV band. Soon after the outburst began, it rapidly transitioned into the soft state. Our goal is to understand how the accretion disk structure deviates from a standard thin disk when the accretion rate is near Eddington. To do so, we analyzed spectra obtained from quasi-simultaneous observations conducted by the Hard X-ray Modulation Telescope (Insight-HXMT), the Nuclear Spectroscopic Telescope Array (NuSTAR), and the Neil Gehrels Swift Observatory (Swift). These spectra are well-fitted by a model comprising a disk, a weak corona, and a reflection component. We suggest that the reflection component is caused by disk self-irradiation, that is by photons emitted from the inner disk which return to the accretion disk surface, as their trajectories are bent by the strong gravity field. In this scenario, the best-fitting parameters imply that the reflected flux represents more than half of the total flux. Using general relativistic ray-tracing simulations, we show that this scenario is viable when the disk becomes geometrically thick, with a funnel-like shape, as the accretion rate is near or above the Eddington limit. In the specific case of 4U 1543-47, an angle $\gtrsim$ 45 deg between the disk surface and the equatorial plane can explain the required amount of self-irradiation.}
  

   \maketitle
%
%________________________________________________________________

\section{Introduction}
A black hole binary (BHB) is a system consisting of a stellar-mass black hole (BH) with strong gravity and a nondegenerate secondary star. These systems are often triggered by accretion instability \citep{2001Lasota}, leading to highly variable behavior. BHBs typically exhibit outbursts that can last for weeks to months, during which the flux can vary by up to seven orders of magnitude \citep[e.g.][]{2006Remillard}. Throughout an outburst, BHBs commonly transit through distinct spectral states \citep[e.g.][]{2004Fender,2005Belloni,2006Remillard}. The X-ray spectra in these states are characterized by the relative contributions of two principal components: thermal and non-thermal components. The hard state is typically observed at the beginning and end of an outburst when the source is relatively faint. In this state, the spectra are dominated by a power-law (PL) component emitted from a corona or the base of a jet, generated through inverse-Compton scattering of thermal seed photons from the disk \citep{2005Markoff}. As the outburst progresses towards its peak, the source enters the soft state. In this state, the X-ray spectra are primarily composed of a multi-temperature disk blackbody, characterized by a typical temperature of around $\sim$1\,keV \citep{1973Shakura}. Moreover, many outbursts also exhibit intermediate states during the transitions between these two states, with the hard-to-soft transitions being brighter than the reverse transitions. Overall, a BHB typically follows a counterclockwise q-shaped trajectory in the hardness-intensity diagram (HID) throughout the course of an outburst \citep[e.g.,][]{Homan2001}.

In addition to the disk and corona components, the X-ray spectra of many BHBs exhibit a notable reflection component \citep{2001Done}. Typically, this reflection component arises when the corona illuminates the accretion disk in the hard and intermediate states \citep{2014Garcia}. The reflection spectra display distinct features, including an iron line at around $\sim$6--7\,keV, which is broadened by the Doppler effect, Compton scattering, and gravitational redshift \citep{2000Fabian}, as well as a Compton hump in the energy range of approximately $\sim$20--40\,keV. However, recent investigations have observed significant reflection components even in soft states \citep{2020Connors,2021Connors,2021Lazar}. The presence of reflection in soft states cannot be adequately explained solely by coronal illumination, as the corona component should be considerably weaker during the soft state. Instead, it is likely that the reflection originates from the returning thermal radiation \citep{1976Cunningham,2021Riaz}, where the disk dominates the emission and its photons are bent back by the strong gravitational pull of the black hole, thereby illuminating the disk surface.

%accretion geometry
The accretion disk in a BHBs exhibits distinct characteristics in different states. In the hard state, the disk is believed to be truncated at a specific radius, gradually extending toward the innermost stable circular orbit (ISCO) as the system transitions to the soft state \citep[e.g.,][]{2004Fender,2006Remillard}. Additionally, when the accretion rate exceeds approximately 30\% of the Eddington limit, the inner disk deviates from a standard thin disk and expands into a geometrically thick disk due to the dominance of radiation pressure in the inner disk region \citep{1988Abramowicz,Abramowicz2005SuperEddingtonBH}. General relativistic radiative magneto-hydrodynamics (GRRMHD) simulations \citep{2022Wielgus,2023Huang} suggest that other features start to become significant in the near-Eddington regime, such as mass-loaded disk outflows, geometric collimation inside a polar funnel, a hot comptonizing region sandwiching the disk, a flatter disk temperature profile and a reduced radiative efficiency from the inner disk. Such features become more prominent in the super-Eddington regime (typical for example of ultraluminous X-ray sources). Phenomenologically, the near-Eddington transitional regime was already noted in early studies of Galactic BH transients \citep{2004Kubota&Makishima}. Thus, nearby BH transients that approach or exceed the Eddington limit during their outburst can provide important information on the physical evolution of an accretion flow from standard disk to ultraluminous regime. The recent outburst of the Galactic BH 4U~1543$-$47 is an excellent case study for such a regime.

%Simultaneously, a hot corona forms above the black hole \citep{2022Wielgus,2023Huang}. These variations in the structure of the accretion disk and the presence of a hot corona are prominent features characterizing different states of BHBs.

%By examining the spectral features, gaining insights into the accretion geometry becomes possible, leading to a deeper understanding of the accretion process. Particularly, when a source is situated within our galaxy and accretes near or surpasses the Eddington limit, it offers a unique advantage compared to distant extragalactic BHBs. Such proximity enhances the potential for revealing intricate details regarding the accretion geometry at high luminosity. The recent 2021 outburst of the Galactic source, 4U 1543--47, presents an excellent opportunity to study and uncover additional insights into the accretion dynamics under these conditions.

4U 1543--47 was first discovered in 1971 \citep{1972Matilsky}. It showed outbursts nearly every 10 years, i.e., in 1971, 1983, 1992 and 2002 \citep{1984Kitamoto,2004Buxton}. After remaining in quiescence for around 20 years, it underwent a new outburst in 2021, which was first captured by Monitor of All-sky X-ray Image \citep[MAXI;][]{2009Matsuoka,2021ATel14701} on June 11th, 2021. Its flux increases quickly since the start of the outburst, and the peak observed flux is up to $ \rm \sim 1.96\times10^{-7}~erg~cm^{-2}~s^{-1}$ ($\sim$8.2\,Crab) in the 2--10\,keV band \citep{2021ATel14708}, much brighter than that of previous outbursts, which are $\sim$1.9\,Crab (2--6\,keV) in 1971 \citep{1972Matilsky}, $\sim$4\,Crab (3.7--7.5\,keV) in 1983 \citep{1984Kitamoto} and $\sim$3.3\,Crab (3.7--7.5\,keV) in 2002 \citep{2004Park,2020Russell}, respectively. This outburst is thus the brightest one that MAXI and the Neil Gehrels Swift Observatory (Swift) have ever observed among BHBs \citep{2021ATel14708,2021ATel14725}. By assuming a distance of 7.5$\pm$1.0\,kpc \citep{2002Orosz,2004Park,2004Jonker}, the peak bolometric luminosity is larger than $\rm 10^{39}~erg~s^{-1}$, approaching the Eddington limit \citep{2021ATel14708} of the BH with mass of 9.4 $\pm$ 2.0 $M_{\odot}$ \citep{orosz2003inventory,2004Park}. 

The orbital inclination of this source is determined to be 20.7$^{\circ}$ $\pm$ 1.5$^{\circ}$ \citep{orosz2003inventory}, while the disk inclination is suggested as $\gtrsim$ 30$^{\circ}$ \citep{2014Morningstar,2020Dong}. There is a discrepancy in spin measurements for this source. \cite{2006Shafee} reported a spin of $0.8\pm 0.1$ using the continuum-fitting method. Subsequently, \cite{2009Miller} and \cite{2014Morningstar} utilized the method of continuum and reflection joint fitting, suggesting a low spin of $0.3\pm 0.1$ and $0.43_{-0.31}^{+0.22}$, respectively. More recently, \cite{2020Dong} constrained the spin to be $0.67_{-0.08}^{+0.15}$ via the reflection fitting.

Some X-ray telescopes, such as the Hard X-ray Modulation Telescope \citep[Insight-HXMT;][]{2020Zhang}, Swift, and the Nuclear Spectroscopic Telescope Array \citep[NuSTAR;][]{2013Harrison}, have monitored the 2021 outburst. Jin et al. 2023 (submitted) provided an analysis of the long-term evolution of light curves and energy spectra using Insight-HXMT data with some pre-fixed parameters.
%, our understanding of the accretion geometry remains limited. 
In this paper, we will present the spectral results from the quasi-simultaneous Insight-HXMT, NuSTAR, and Swift observations by freeing these parameters, and focus on the accretion geometry at high luminosity. The observational details and data analysis procedures are described in Section~\ref{sec:obs}, followed by the presentation of our spectral results in Section~\ref{sec:res}. Finally, in Section~\ref{sec:dis}, we discuss our findings.

%__________________________________________________________________

\section{observations and data reduction}
\label{sec:obs}

\begin{figure*}
    \centering
    \includegraphics[width=1\textwidth]{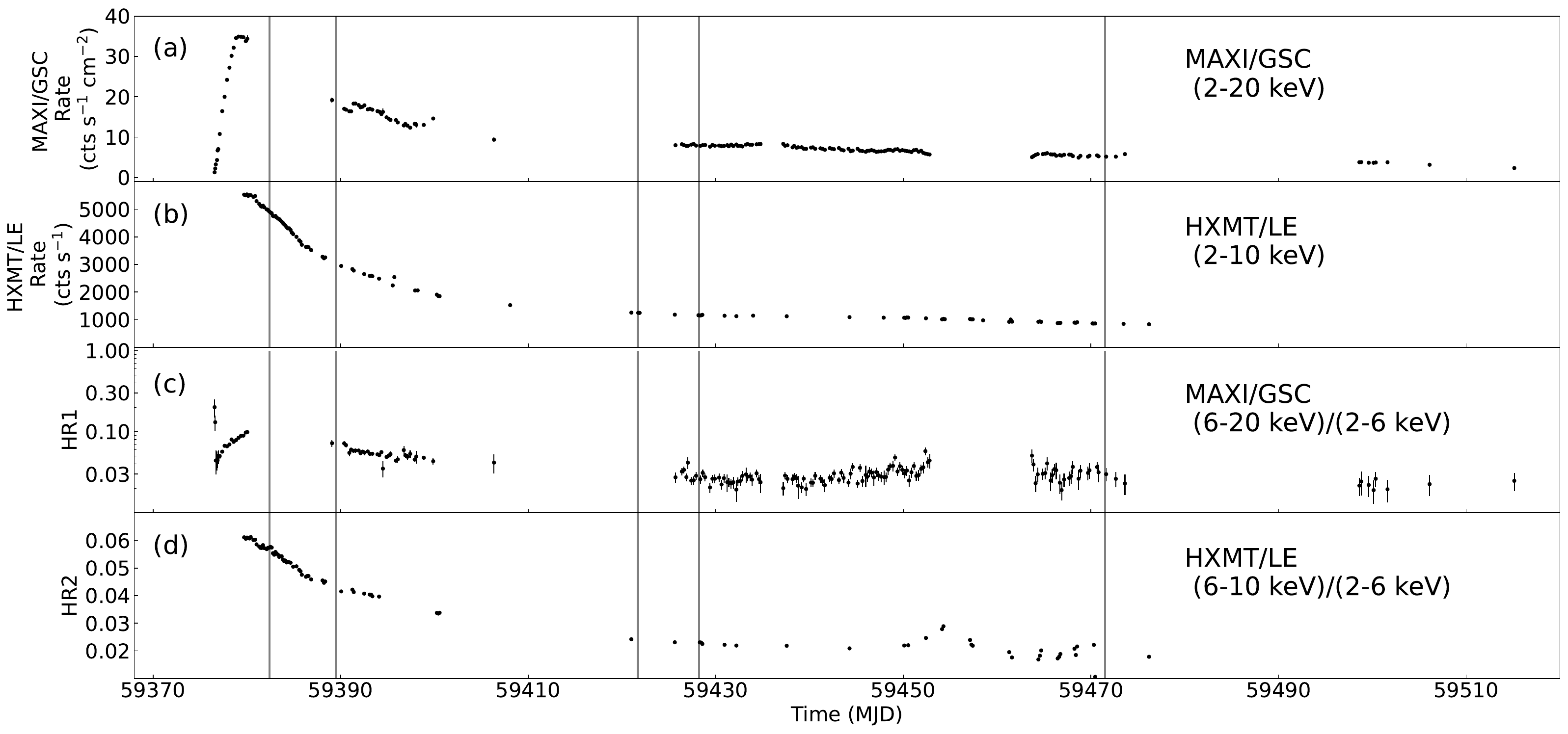}
    \caption{The evolution of count rates and HRs with time. (a) MAXI/GSC 2--20 keV light-curve. (b) Insight-HXMT/LE 2--10 keV light-curve. (c) MAXI/GSC HR defined as the ratio of (6--20 keV)/(2-–6 keV) count rates. (d) Insight-HXMT/LE HR defined as the ratio of (6--10 keV)/(2–-6 keV) count rates. Vertical lines indicate five quasi-simultaneous data sets of Insight-HXMT, NuSTAR, and Swift. }
    \label{fig:lc}
\end{figure*}

\begin{figure*}
    \centering

    \includegraphics[width=1\textwidth]{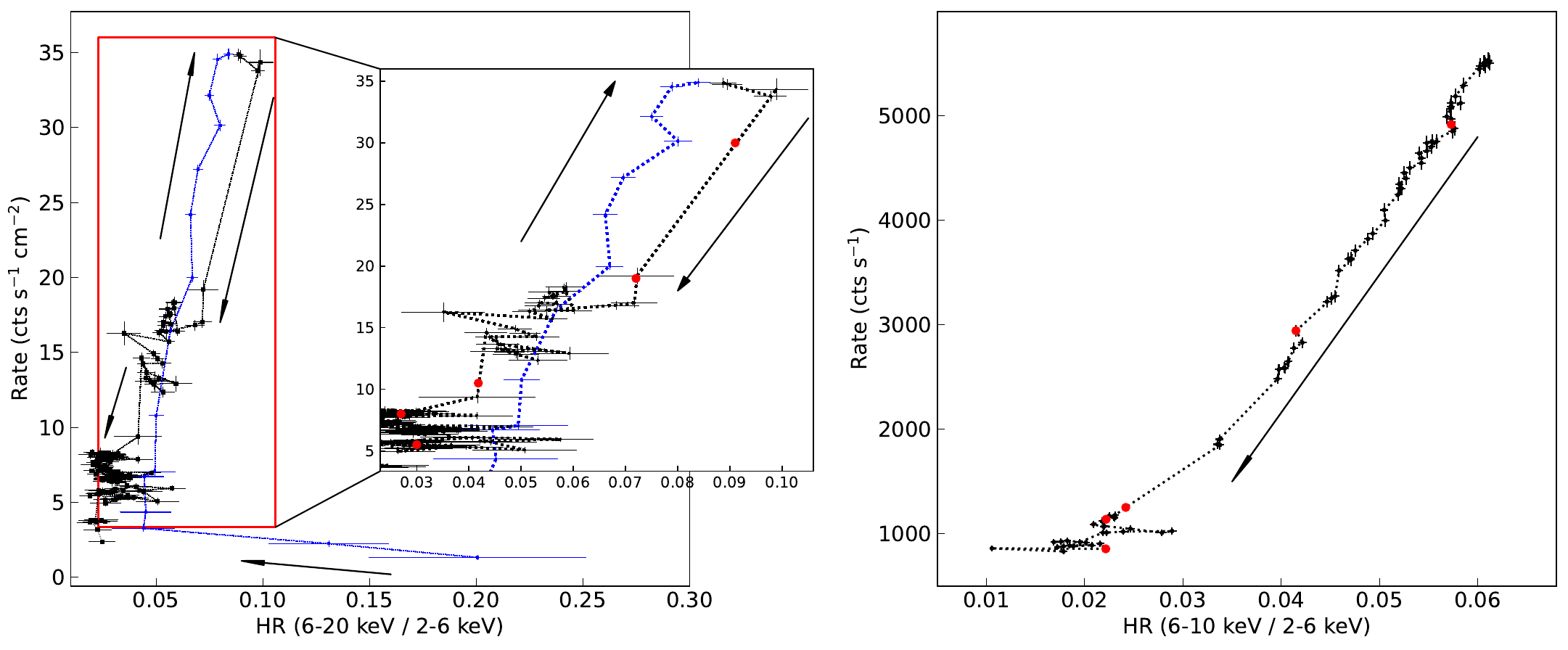}
    \\
    \caption{Hardness-intensity diagrams (HIDs) from MAXI/GSC (left panel) and Insight-HXMT/LE (right panel). The blue dotted line represents the rising phase before the outburst peak, while the black one shows the decay phase. The red dots represent the five data sets used in this paper. If there are no simultaneous MAXI data for some data sets, the MAXI data in approximate date are marked. }
    \label{fig:hid}
\end{figure*}
%%%%%%%%%%%%%%%%%
%%%%%%%%%%%%%%%%%%%%%%%%%%%%%%%%%%%%%%%%%%%%%%%%%%%%%%%%%%%%%%%
\begin{table*}
\begin{threeparttable}
\centering
\caption{Observation information of NuSTAR, Swift/XRT and Insight-HXMT}
\label{tab:obs}
\begin{tabularx}{\textwidth}{@{}>{\centering\arraybackslash}X>{\centering\arraybackslash}X>{\centering\arraybackslash}X>{\centering\arraybackslash}X>{\centering\arraybackslash}X@{}}
\hline
\hline
Mission & ObsID &  Start Time &  End Time & Exposure (s) \\
\hline
\multicolumn{5}{c}{Data Set 1}                                              \\
NuSTAR  & 80702317002 & 2021 Jun 17 10:11:09 & 2021 Jun 17 19:21:09 & 1335.0                        \\
Swift/XRT  & 00014374005 & 2021 Jun 16 19:14:07 & 2021 Jun 16 19:29:55 & 944.7                         \\ 
Insight-HXMT    & P0304026001 & 2021 Jun 17 02:55:11 & 2021 Jun 17 22:00:48 & 11250.0                        \\
%\hline
\hline
\multicolumn{5}{c}{Data Set 2}                                                                         \\
NuSTAR  & 80702317004 & 2021 Jun 24 11:21:09 & 2021 Jun 24 22:46:09 & 3113.1                             \\
Swift/XRT   & 00014374006 & 2021 Jun 24 09:10:53 & 2021 Jun 24 23:29:56 & 559.6                          \\ 
Insight-HXMT    & P0304026005 & 2021 Jun 25 01:27:22 & 2021 Jun 25 06:18:41 & 2876.0                         \\ 
 %\hline
 \hline
\multicolumn{5}{c}{Data Set 3}                                                                         \\
NuSTAR  & 90702326002 & 2021 Jul 26 16:11:09 & 2021 Jul 27 03:06:09 & 5717.3                           \\
Swift/XRT  & 00089352001 & 2021 Jul 27 00:28:19 & 2021 Jul 27 02:19:56 & 1828.2                         \\ 
Insight-HXMT    & P0304026019 & 2021 Jul 26 17:23:25 & 2021 Jul 27 00:16:16 & 1477.0                          \\ 
%\hline
\hline
\multicolumn{5}{c}{Data Set 4}                                                                         \\
NuSTAR  & 90702326004 & 2021 Aug 02 04:21:09 & 2021 Aug 02 15:36:09 & 8024.4                             \\
Swift/XRT   & 00089352002 & 2021 Aug 02 12:37:34 & 2021 Aug 02 13:05:55 & 1698.9                            \\
Insight-HXMT    & P0304026022 & 2021 Aug 02 03:36:12 & 2021 Aug 02 16:24:11 & 3659.0                            \\ 
%\hline
\hline
\multicolumn{5}{c}{Data Set 5}                                                                         \\
NuSTAR  & 90702326012 & 2021 Sep 14 12:26:09 & 2021 Sep 14 21:51:09 & 7409.0                             \\
Swift/XRT   & 00089352004 & 2021 Sep 14 14:29:57 & 2021 Sep 14 14:56:56 & 1619.1                            \\
Insight-HXMT    & P0304026039 & 2021 Sep 13 03:54:01 & 2021 Sep 13 13:31:24 & 3499.0                      \\
\hline

\end{tabularx}
\begin{tablenotes} 
\item Notes. For NuSTAR and Insight-HXMT, the observation logs of FPMA and LE are listed as a representation, respectively.
\end{tablenotes} 
\end{threeparttable}
\end{table*}
%%%%%%%%%%%%%%%%%%%

Insight-HXMT observed 4U 1543--47 from June 14, 2021, to September 19, 2021, accumulating a total of 52 observations. Additionally, there are 10 observations conducted by NuSTAR and 12 observations by Swift/XRT during the outburst. For this study, we utilize five data sets obtained from quasi-simultaneous observations of Insight-HXMT, NuSTAR, and Swift/XRT, covering the period from June 17th to September 14th, 2021. Detailed information regarding these data sets is provided in Table~\ref{tab:obs}. From the light curves of MAXI\footnote{The 2--20\,keV count rates of MAXI are obtained in \url{http://134.160.243.88/top/index.html}.} and Insight-HXMT (Figure~\ref{fig:lc}), we can see that the observations are taken in the outburst peak and decay phase, and cover the typical evolution stages of the complete outburst. The total exposure of Insight-HXMT, NuSTAR, and Swift/XRT are 22761\,ks, 25599\,ks, and 6651\,ks, respectively.

\subsection{Insight-HXMT}
\label{sec:hxmt}
We perform data reduction using the Insight-HXMT Data Analysis Software ({\tt HXMTDAS} {\tt v2.05}\footnote{\url{http://hxmten.ihep.ac.cn/software.jhtml}}) and the latest calibration database files ({\tt CALDB v2.06}). To select the good time intervals, we apply the following criteria: (1) Earth elevation angle $>$10$^{\circ}$; (2) pointing offset angle $<$0.04$^{\circ}$; (3) geomagnetic cutoff rigidity $>$ 8\,GV; (4) at least 300\,s away from the crossing of the South Atlantic Anomaly region. The background of the Low Energy (LE), Medium Energy (ME), and High Energy (HE) telescopes are respectively created with the scripts {\tt lebkgmap}, {\tt mebkgmap}, and {\tt hebkgmap}, based on the Insight-HXMT background models \citep{Liao2020a, Guo2020, Liao2020b}. The response files of LE, ME, and HE are generated by {\tt lerspgen}, {\tt merspgen}, and {\tt herspgen}, respectively. 

Each observation consists of a few individual 
sub-exposures. To improve the statistical accuracy, we combine the spectra of all sub-exposures from each observation. Following the recommendation of the Insight-HXMT calibration group, the combined spectra are rebinned as follows: (1) LE: channels 0 through 579 and 580 through 1535 are respectively grouped with 3 and 5 bins in each group; (2) ME: channels 0 through 1023 are grouped with 6 bins in each group; (3) HE: channels 0 through 255 is grouped with 16 bins in each group. Additionally, systematic errors of 0.5\%, 1\%, and 1\% are applied to the LE, ME, and HE spectra, respectively.

\subsection{NuSTAR}
The {\tt nupipeline} routine of {\tt NuSTARDAS v1.9.7} in {\tt HEASoft v6.29} with {\tt CALDB v20211115}, is employed to process the cleaned event files. Due to high brightness, some source counts may be erroneously vetoed by a noise filter in the {\tt nupipeline} tool, resulting in a lower measured source flux and a discrepancy between the two focal plane modules (FPMA and FPMB)\footnote{\url{https://heasarc.gsfc.nasa.gov/docs/nustar/analysis/}}. To address this issue, following the recommendation provided in the analysis guide, we modify the {\tt statusexpr} keyword in the {\tt nupipeline} tool using the expression "STATUS==b0000xxx00xxxx000". This modification adjusts the behavior of the {\tt nufilter} tool accordingly. The {\tt nuproducts} tool is then used to extract the source events, by adopting a circular region surrounding the source with a radius of $100^{\prime \prime}$ to optimize the signal-to-noise ratio of the spectra \footnote{The script can be found in \url{https://github.com/NuSTAR/nustar-gen-utils/tree/3a603ca820a93c81414a298fd90d2e5a05f5e24a/notebooks}.}. The corresponding background extraction region is a nearby source-free circle with a radius of $100^{\prime \prime}$. The spectra are rebinned with 50 counts per bin at least.

\subsection{Swift}
The Swift/XRT spectra are extracted through the Swift-XRT data products generator \citep{Evans2009}\footnote{\url{https://www.swift.ac.uk/user\_objects/}} using {\tt HEASoft v6.29}. The best source position is determined by enabling the centroiding option, using a single-pass cell-detect method and a search radius of 1$^{\prime}$. Moreover, a maximum of 10 individual observations are allowed to do source detection before using the stacked image. Since the source is very bright, in order to eliminate the pile-up effect to the greatest extent, we only select the grade 0 events in windowed timing mode, and adjust the inner radius of the annulus source region (outer radius is 20 pixels) to prevent the count rate from exceeding 150 ct~s$^{-1}$, as recommended by the Swift help desk. The spectra are rebinned with 25 counts per bin at least. 

%__________________________________________________________________

\section{Analysis and results}
\label{sec:res}

\begin{figure}
    \centering
    \includegraphics[scale=0.325]{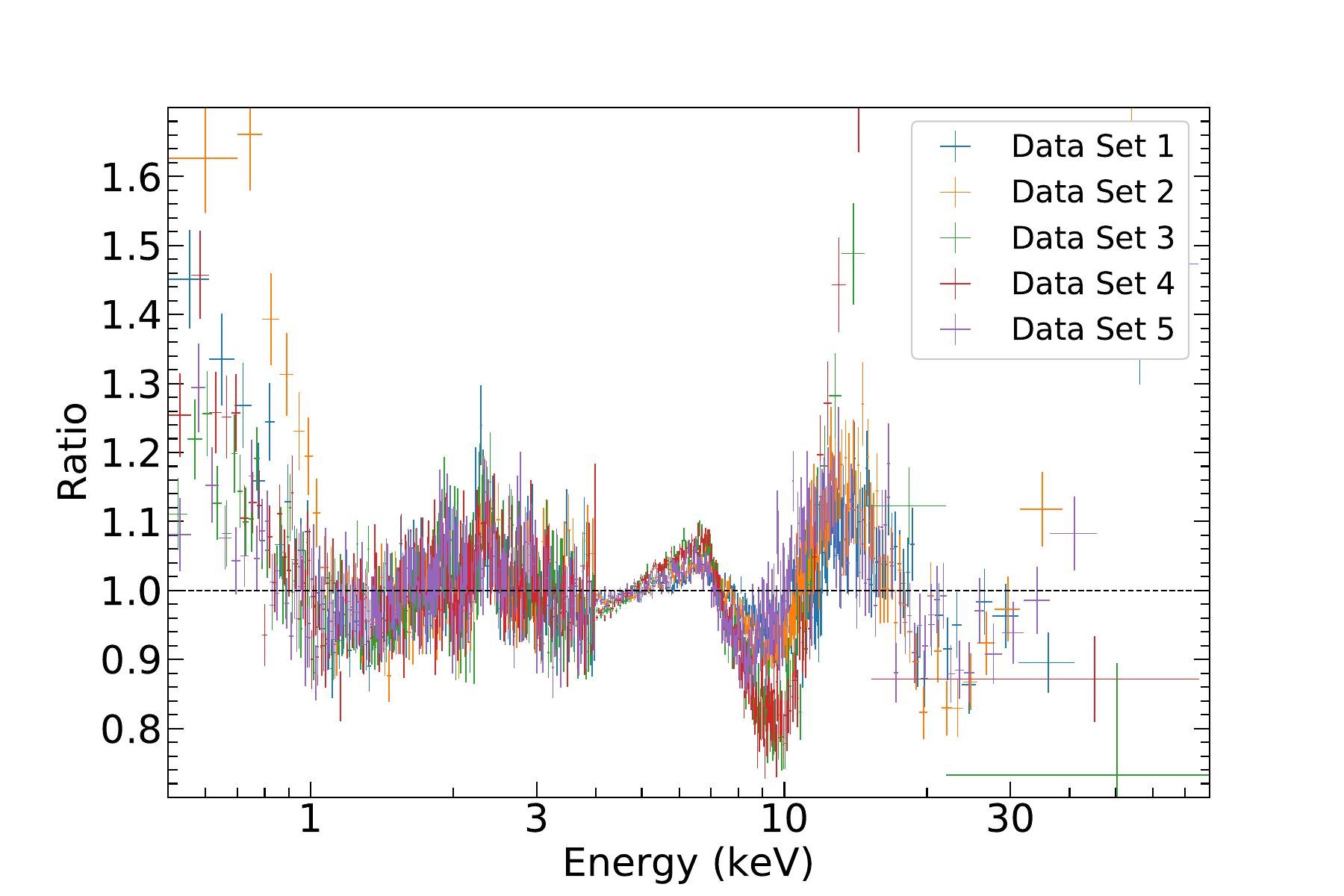}\\
    \caption{Data-to-model ratios of {\tt model 2} ({\tt tbabs*(diskbb+nthcomp)}) for Swift/XRT and NuSTAR/FPMA. Different data sets are marked in different colors. The plots are grouped to have a S/N $\ge$ 20 per bin for display clarity. \\}
 \label{fig:ratio}
   
\end{figure}

\begin{figure}
    \centering
    \includegraphics[scale=0.63]{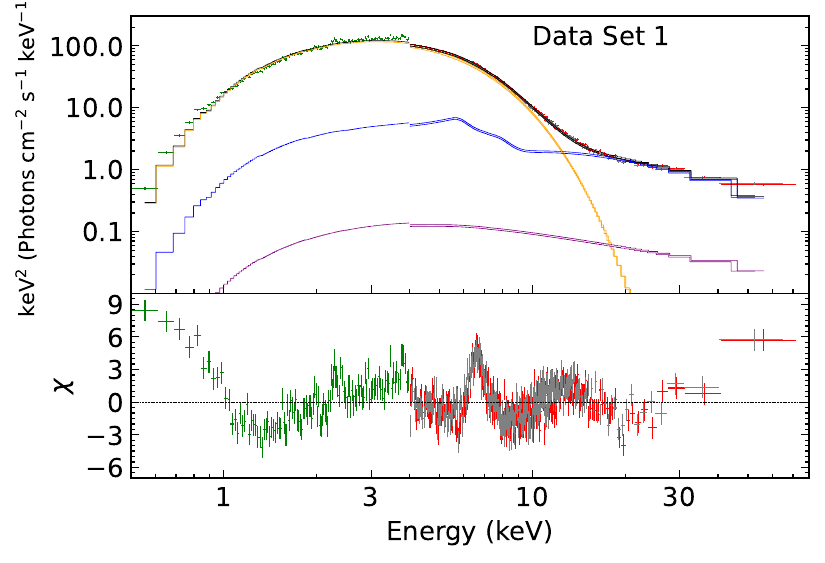}\\
    \caption{Spectra, model components, and spectral residuals for {\tt tbabs*(diskbb+nthcomp+relxillcp*nthratio)}. Data Set 1 of Swift and NuSTAR are taken as a representative. The green, red, and gray data points show the Swift, and NuSTAR/FPMA and FPMB data, respectively.  The black solid line is the total model, and the orange, purple, and blue solid lines show the {\tt diskbb}, {\tt nthcomp} and {\tt relxillcp*{\tt nthratio}} components, respectively.\\}
    \label{fig:relxillcp}
\end{figure}

\begin{figure}
    \centering
    \includegraphics[scale=0.5]{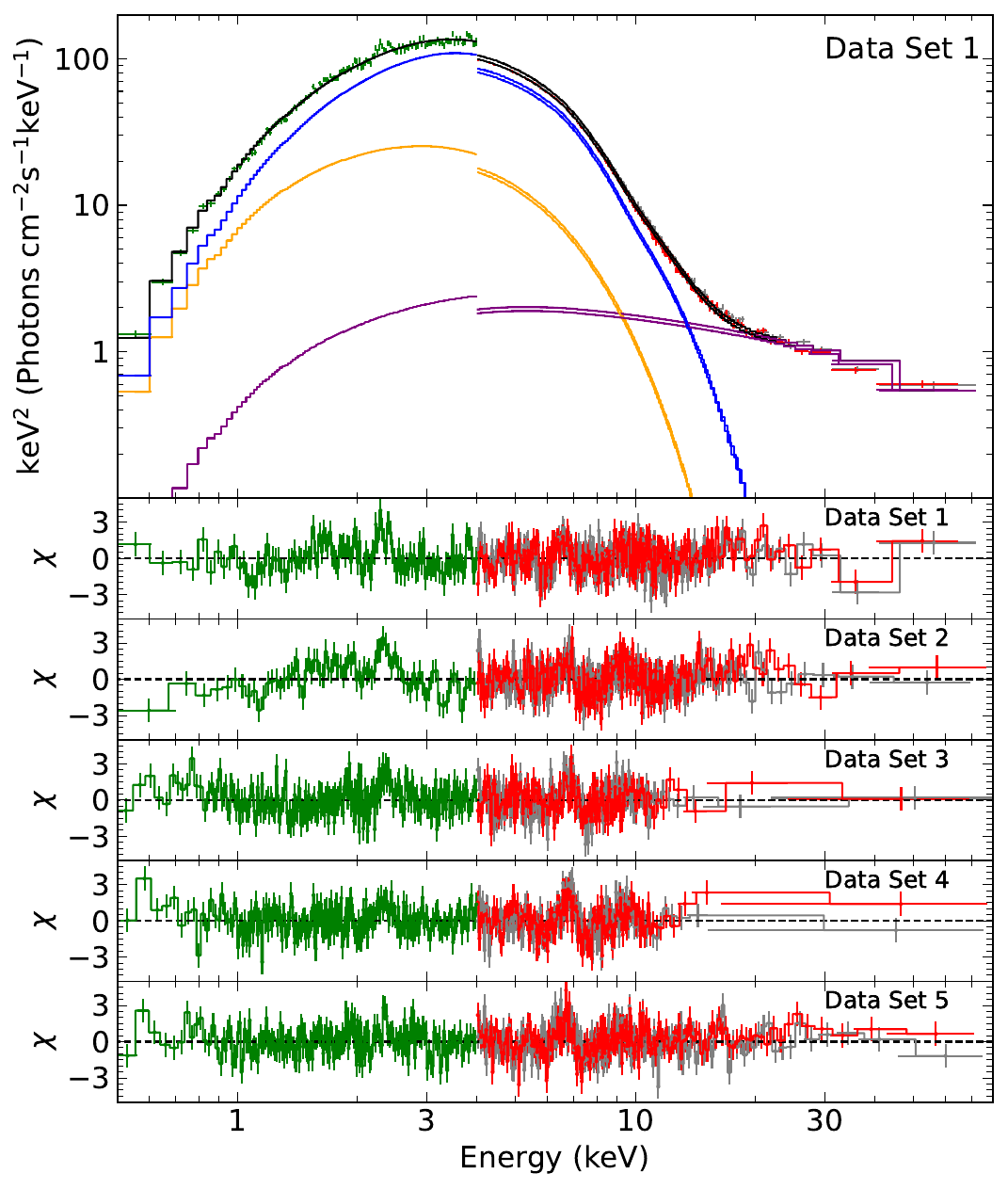} \\
    \caption{Spectra, model components, and spectral residuals for Swift/XRT (green) and NuSTAR (red and gray) with {\tt model 3} ({\tt tbabs*(diskbb+nthcomp+relxillNS)}). The black solid line is the total model, and the orange, purple, and blue solid lines show the {\tt diskbb}, {\tt nthcomp}, and {\tt relxillNS} components, respectively.\\}
    \label{fig:NS_wtga}
\end{figure}

\begin{figure}
    \centering
    \includegraphics[scale=0.48]{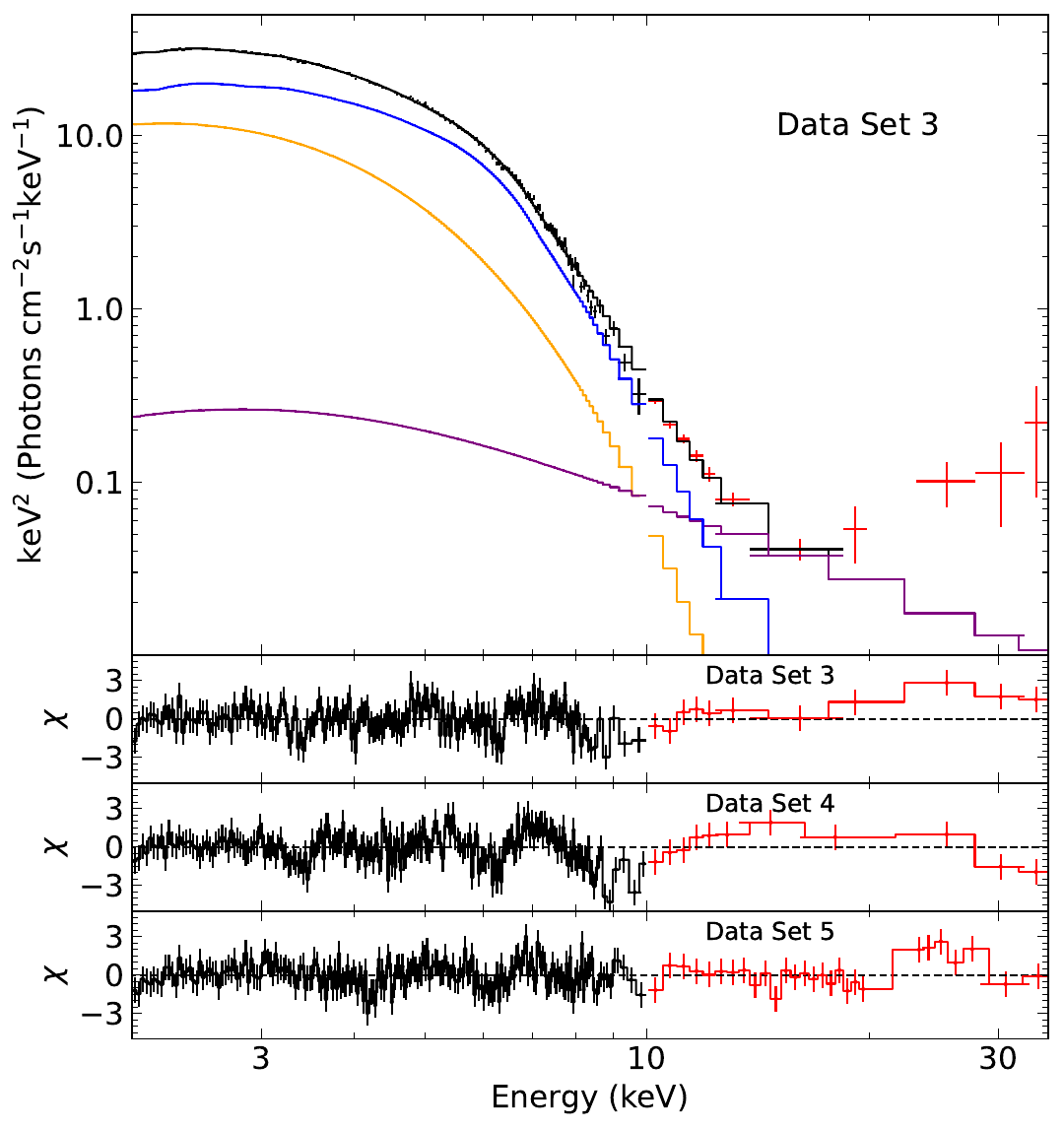}\\
    \caption{Spectra (black for LE and red for ME), model components, and spectral residuals for Insight-HXMT with {\tt model 3}. The black solid line is the total model fitted to the data, and the orange, purple and blue solid lines show the {\tt diskbb}, {\tt nthcomp} and {\tt relxillNS} spectral components, respectively.}
    \label{fig:hxmt_cp_ns}
\end{figure}

%%%%%%%%%%%%%%%%%%%%%%%%%%%%%%%%%%%%%%%%%%%%%%%%%%%%%%%%%%%%%%%%%%%%%%%%%%%%%%%%%%%%%%%%%%
\begin{table*}
\begin{threeparttable}
\centering
\renewcommand{\arraystretch}{1.3}
\caption{Joint fitting parameters of Swift/XRT and NuSTAR for {\tt model 3} ({\tt tbabs*(diskbb+nthcomp+relxillNS)}).}
\label{tab:sw_nu_dbb_ilink}
\begin{tabularx}{\textwidth}{@{}>{\centering\arraybackslash}X>{\centering\arraybackslash}X>{\centering\arraybackslash}X>{\centering\arraybackslash}X>{\centering\arraybackslash}X>{\centering\arraybackslash}X@{}}
\hline\hline
Parameter & Data Set 1 & Data Set 2 & Data Set 3 & Data Set 4 & Data Set 5 \\
\hline
\multicolumn{6}{c}{\tt tbabs} \\
$N_{\rm H}^{\bigstar}\ \rm(10^{22}\ cm^{-2})$
  &  \multicolumn{5}{c}{$0.450\pm 0.009$}
\\
\hline
\multicolumn{6}{c}{\tt diskbb} \\
$T_{\rm in}\ \rm (keV)$
  &  $1.162_{-0.010}^{+0.008}$
  &  $1.076_{-0.007}^{+0.006}$
  &  $0.915\pm0.008$
  &  $0.907\pm0.007$
  &  $0.872\pm0.008$
  \\
norm ($10^3$)
  &  $2.4_{-0.4}^{+0.3}$
  &  $2.2\pm0.3$
  &  $2.5\pm0.3$
  &  $2.1_{-0.3}^{+0.2}$
  &  $2.7\pm0.2$
  \\
\hline
\multicolumn{6}{c}{\tt nthcomp} \\
$\rm \Gamma$
  &  $2.35_{-0.08}^{+0.11}$
  &  $2.09_{-0.08}^{+0.09}$
  &  $3.4_{-0.6}^{+0.2}$
  &  $3.4_{-0.4}^{+0.3}$
  &  $2.02\pm0.04$
\\
$kT_{\rm e}$ (keV)
  &  $21.7_{-6.0}^{+19.6}$
  &  $16_{-3}^{+6}$
  &  $>7.9$
  &  $>14.2$
  &  $30.5_{-8.7}^{+26.6}$
 \\
norm
  &  $0.77_{-0.10}^{+0.13}$
  &  $0.218_{-0.03}^{+0.04}$
  &  $0.09_{-0.04}^{+0.03}$
  &  $0.061_{-0.019}^{+0.025}$
  &  $0.129\pm0.011$
 \\
\hline
\multicolumn{6}{c}{\tt relxillNS} \\
$q_{\rm in}$ ($=q_{\rm out}$)
  &  $3.0\pm0.2$
  &  $3.2\pm0.2$
  &  $4.2_{-0.3}^{+0.4}$
  &  $4.3_{-0.3}^{+0.4}$
  &  $4.3_{-0.5}^{+0.7}$
  \\
$i^{\bigstar}$ ($^{\circ}$)
  &  \multicolumn{5}{c}{$32.2_{-1.2}^{+2.3}$}  
  \\
$R_{\rm in}\ (R_{\rm ISCO})$
  &  $<1.56$
  &  $<1.27$
  &  $<1.11$
  &  $<1.11$
  &  $<1.16$
  \\
log~$\xi~\rm{(erg~cm~s^{-1})}$
  &  $>3.91$
  &  $3.94_{-0.05}^{+0.04}$
  &  $3.65\pm0.08$
  &  $3.79\pm0.10$
  &  $3.72_{-0.09}^{+0.11}$
  \\
$A_{\rm Fe}^{\bigstar}~\rm{(solar)}$
  &  \multicolumn{5}{c}{$2.7\pm0.3$}  
  \\
log~$N$ ($\rm cm^{-3}$)
  &  $>18.8$
  &  $18.5\pm0.2$
  &  $17.9_{-0.9}^{+0.5}$
  &  $17.1_{-1.5}^{+0.9}$
  &  $<17.6$
  \\
norm ($10^{-3}$)
  &  $204_{-13}^{+19}$  &  $132_{-9}^{+11}$  &  $66\pm5$  &  $64_{-2}^{+5}$  &  $40_{-3}^{+3}$
\\
\hline
$F_{\rm abs}$ (0.5--10\,keV)
& 24.0 & 14.9 & 6.80 & 6.13 & 4.78  \\
$F_{\rm abs}$ (0.5--100\,keV)
& 24.6 & 15.1 & 6.81 & 6.14 & 4.88 \\
$F_{\rm unabs}$ (0.5--100\,keV)
& 28.6 & 17.9 & 8.69 & 7.76 & 6.31 \\
$L_{\rm X}/L_{\rm Edd}$
& 1.64 & 1.03 & 0.50 & 0.44 & 0.36 \\
\hline
$F_{\rm relxillNS}/ F_{\rm X}$
& 0.67 & 0.67 & 0.60 & 0.65 & 0.50 \\
\hline
$\chi^2$/dof
& \multicolumn{5}{c}{6284.0/5276 = 1.19}    \\
\hline

\end{tabularx}
\begin{tablenotes} 
\item Notes. All errors are quoted at the 90\% confidence level. Taking Data Set 3 as an example, the probability distributions of parameters obtained through the Markov chain Monte Carlo (MCMC) algorithm are presented in Figure~\ref{fig:corner1}; 

\item $\bigstar$ indicates that the parameters between different data sets are linked;
\item $N_{\rm H}$ is the X-ray absorption column density in units of $10^{22}$~atoms~$\rm cm^{-2}$;

 \item $T_{\rm in}$ is the inner disk temperature in units of keV;
 \item $\Gamma$ is the power-law photon index of the thermal Comptonization component;
 \item $kT_{\rm e}$ is the electron temperature of the corona in units of keV, standing for the high energy cutoff of the Comptonization continuum;

\item $q_{\rm in}$ and $q_{\rm out}$ are respectively the inner and outer emissivity indices for the coronal flavor models;
\item $R_{\rm in}\ (R_{\rm ISCO})$ is the inner radius of the accretion disk in units of $R_{\rm ISCO}$;
\item $A_{\rm Fe}$ is the iron abundance of the accretion disk in units of solar abundance;
\item $\xi$ is the ionization parameter of the accretion disk in units of $\rm erg~cm~s^{-1}$;
\item $N$ is the density of the accretion disk in units of $\rm cm^{-3}$;
\item $i$ is the disk inclination angle in units of deg;
\item $\rm norm$ is the normalization;
\item $F_{\rm abs}$ and $F_{\rm unabs}$ are absorbed and unabsorbed fluxes in units of $\rm 10^{-8}~erg~cm^{-2}~s^{-1}$ ;
\item $L_{\rm X}$ is the absorption-corrected 0.5--100\,keV luminosity, calculated by $L_{\rm X}=4 \times \pi \times D^2 \times F_{\rm unabs}$, where D = 7.5\,kpc;
\item $L_{\rm Edd}$ is the Eddington luminosity with BH mass 9.4 $M_{\odot}$, i.e., $L_{\rm Edd}$ = $\rm{1.2\times 10^{39}~erg~s^{-1}}$; 
\item $F_{\rm relxillNS}/ F_{\rm X}$ is the ratio of the flux of the reflection component to the total flux in 0.5--100\,keV.
\end{tablenotes} 
\end{threeparttable}
\end{table*}
%%%%%%%%%%%%%%%%%%%%%%%%%%%%%%%%%%%%%%%%%%%%%%%%%%%%%%%%%%%%%%%%%%
\begin{table*}
\begin{threeparttable}
 \renewcommand\arraystretch{1.3}
 
 \centering
 \caption{Joint fitting parameters of Insight-HXMT for {\tt model 3}. }
 \label{tab:hxmt_dbb_ilink}
 \begin{tabularx}{\textwidth}{@{}>{\centering\arraybackslash}X>{\centering\arraybackslash}X>{\centering\arraybackslash}X>{\centering\arraybackslash}X@{}}
 \hline
 \hline
 Parameter &    Data Set 3 &  Data Set 4 & Data Set 5   \\
 \hline
 \multicolumn{4}{c}{\tt diskbb} \\
$T_{\rm in}\ \rm (keV)$
  &  $0.832_{-0.008}^{+0.009}$
  &  $0.832_{-0.008}^{+0.010}$
  &  $0.816_{-0.007}^{+0.009}$ \\
norm ($10^3$)
  &  4 (fixed) & 4 (fixed) &  $3.9_{-0.4}^{+1.3}$
\\
 \hline

\multicolumn{4}{c}{\tt nthcomp}  \\
$kT_{\rm e}$ (keV)
& unconstrained
& unconstrained
& $>51.7$
  \\
norm
  &  $0.17_{-0.04}^{+0.05}$
  &  $0.12\pm0.03$
  &  $0.165_{-0.009}^{+0.011}$
  \\
 \hline
 
\multicolumn{4}{c}{\tt relxillNS} \\
$q_{\rm in}$ ($=q_{\rm out}$)
&  $5.2_{-0.3}^{+0.4}$
  &  $5.3_{-0.3}^{+0.4}$
  &  $5.1\pm0.4$
\\
$i^{\bigstar}$ ($^{\circ}$)
  & \multicolumn{3}{c}{$34.7_{-0.8}^{+0.7}$}
  \\
log~$\xi~\rm{(erg~cm~s^{-1})}$
 &  $2.31\pm0.03$
  &  $2.29_{-0.04}^{+0.02}$
  &  $2.24_{-0.03}^{+0.09}$
\\
$A_{\rm Fe}^{\bigstar}~\rm{(solar)}$
 & \multicolumn{3}{c}{$2.2_{-0.3}^{+0.4}$}
 \\
log~$N$ ($\rm cm^{-3}$)
 &  $17.2_{-0.4}^{+0.5}$
  &  $17.8_{-0.3}^{+0.6}$
  &  $>17.4$
\\
norm ($10^{-3}$)
  &  $294_{-29}^{+27}$
  &  $236_{-24}^{+20}$
  &  $110\pm13$
 \\ \hline
  $F_{\rm abs}$ (0.5--10\,keV) 
 & 7.77
 & 7.21
 & 5.24
  \\
 $F_{\rm abs}$ (0.5--100\,keV) 
 & 7.78
 & 7.22
 & 5.38
 \\
 $F_{\rm unabs}$ (0.5--100\,keV) 
 & 11.0
 & 10.3
 & 7.92
   \\
 $L_{\rm X}/L_{\rm Edd}$ 
 &0.63
 &0.59
 &0.45  \\ 
 \hline
 $F_{\rm relxillNS}/ F_{\rm X}$
& 0.68 & 0.65 & 0.59 \\ \hline
 $\chi^2$/dof
 &  \multicolumn{3}{c}{910.8/819 = 1.11}
 \\
\hline

%\multicolumn{6}{l}{Notes. $\bigstar$ indicates that the parameters are fixed.}
\end{tabularx}
\begin{tablenotes} 
\item Notes. All errors are quoted at the 90\% confidence level. Taking Data Set 3 as an example, the probability distributions of parameters obtained through MCMC are presented in Figure~\ref{fig:corner2}. 
\end{tablenotes} 
\end{threeparttable}
\end{table*}

%%%%%%%%%%%%%%%%%%%%%%%%%%%%%%%%%%%%%%%%%%%%%%%%%%%%%%%%%%%%%
\begin{table*}
\begin{threeparttable}
\centering
\renewcommand{\arraystretch}{1.3}
\caption{Joint fitting parameters of Swift/XRT and NuSTAR for {\tt model 4} ({\tt tbabs*(slimbh+nthcomp+relxillNS)}). }
\label{tab:sw_nu_slimbh_ilink}
\begin{tabularx}{\textwidth}{@{}>{\centering\arraybackslash}X>{\centering\arraybackslash}X>{\centering\arraybackslash}X>{\centering\arraybackslash}X>{\centering\arraybackslash}X@{}}
\hline\hline
Parameter & Data Set 2 & Data Set 3 & Data Set 4 & Data Set 5 \\
\hline
\multicolumn{5}{c}{\tt tbabs} \\
$N_{\rm H}^{\bigstar}\ \rm(10^{22}\ cm^{-2})$
  &  \multicolumn{4}{c}{$0.436_{-0.010}^{+0.008}$}
\\
\hline
\multicolumn{5}{c}{\tt slimbh} \\
$\rm norm$
  &  $0.32\pm0.03$
  &  $0.30_{-0.02}^{+0.04}$
  &  $0.35\pm0.03$
  &  $0.43_{-0.03}^{+0.02}$
  \\

\hline
\multicolumn{5}{c}{\tt nthcomp} \\
$\rm \Gamma$
  &  $2.11\pm0.09$
  &  $3.0_{-0.7}^{+0.4}$
  &  $3.5_{-0.3}^{+0.2}$
  &  $2.02_{-0.03}^{+0.04}$
\\
$kT_{\rm e}$ (keV)
  &  $17_{-3}^{+6}$
  &  unconstrained
  &  $>17.2$
  &  $30_{-7}^{+21}$
 \\
norm
  &  $0.24\pm0.04$
  &  $0.05\pm0.03$
  &  $0.068_{-0.018}^{+0.026}$
  &  $0.128_{-0.007}^{+0.008}$
 \\
\hline
\multicolumn{5}{c}{\tt relxillNS} \\
$q_{\rm in}$ ($=q_{\rm out}$)
  &  $3.567_{-0.013}^{+0.023}$
  &  $5.034_{-0.038}^{+0.017}$
  &  $5.43\pm 0.02$
  &  $6.43_{-0.54}^{+0.07}$
  \\
$i^{\bigstar}$ ($^{\circ}$)
  &  \multicolumn{4}{c}{$34.5_{-0.18}^{+0.08}$}  
  \\
$R_{\rm in}\ (R_{\rm ISCO})$
  &  $1.35^{+0.15}_{-0.18}$
  &  $<1.01$
  &  $<1.01$
  &  $<1.08$
  \\
$kT_{\rm bb}$
  &  $1.036\pm0.006$
  &  $0.885_{-0.007}^{+0.009}$
  &  $0.913_{-0.006}^{+0.007}$
  &  $0.876\pm0.009$
  \\
log~$\xi~\rm{(erg~cm~s^{-1})}$
  &  $>3.96$
  &  $3.70_{-0.03}^{+0.05}$
  &  $>3.97$
  &  $>3.89$
  \\
$A_{\rm Fe}^{\bigstar}~\rm{(solar)}$
  &  \multicolumn{4}{c}{$2.91_{-0.20}^{+0.14}$}  
  \\
log~$N$ ($\rm cm^{-3}$)
  &  $17.97_{-0.27}^{+0.10}$
  &  $17.1_{-0.9}^{+0.8}$
  &  $<16.7$
  &  $<17.1$
  \\
norm ($10^{-3}$)
  &  $147\pm6$  &  $87\pm3$  &  $0.071\pm0.002$  &  $47_{-2}^{+4}$
  \\
\hline
$F_{\rm abs}$ (0.5--10\,keV)
&  15.0 & 6.86 & 6.18 & 4.83 \\
$F_{\rm abs}$ (0.5--100\,keV)
& 15.2 & 6.87 & 6.19 & 4.93 \\
$F_{\rm unabs}$ (0.5--100\,keV)
& 17.9 & 8.68 & 8.75 & 6.32 \\
$L_{\rm X}/L_{\rm Edd}$
& 1.03 & 0.50 & 0.44 & 0.36 \\
\hline
$F_{\rm relxillNS}/ F_{\rm X}$
& 0.70 & 0.73 & 0.70 & 0.56 \\
\hline
$\chi^2$/dof
& \multicolumn{4}{c}{4948.7/4073=1.22}    \\
\hline 
\end{tabularx}
\begin{tablenotes}
\item Notes. All errors are quoted at the 90\% confidence level. Taking Data Set 3 as an example, the probability distributions of parameters obtained obtained through MCMC are presented in Figure~\ref{fig:corner3}.
\end{tablenotes} 
\end{threeparttable}
\end{table*}
%%%%%%%%%%%%%%%%%%%%%%%%%%%%%%%%%%%%%%%%%%%%%%%%%%%%%%%%%%%%%%%%%%

Before conducting a detailed spectral analysis, we initially examine the spectral states of the five data sets using the hardness ratio (HR) and HID. The HRs are calculated as the ratio of the 6--20\,keV count rates to the 2--6\,keV count rates from MAXI (HR1). Data points with a signal-to-noise ratio below 3$\sigma$ are excluded. In Panel c of Figure~\ref{fig:lc}, we observe that HR1 rapidly decreases to 0.04--0.05 at the beginning of the outburst. It then gradually increases before displaying a long-term decay. Panel d shows HR2, which is obtained by calculating the ratio of the 6--10\,keV count rates to the 2--6\,keV count rates from Insight-HXMT. Due to the late start of the observations, the HR has already begun to decline. Combining the intensity and HR, as shown in Figure~\ref{fig:lc}, we construct the HID using data from MAXI/GSC and Insight-HXMT/LE, as illustrated in Figure~\ref{fig:hid}. The black arrows in the figures represent the evolutionary direction. From the left panel, it is evident that the system does not follow the typical counterclockwise q-shaped track observed in BHBs. Instead, it transitions to the soft state shortly after the onset of the outburst. Although the subsequent HRs are positively correlated with the count rates, the HRs do not exceed 0.1. The red data points indicate the five data sets used in this study, and they all correspond to a very soft state.

In order to fully understand the spectral behaviors over a broad energy band, we perform joint modeling of the spectra from NuSTAR and Swift/XRT. However, to mitigate calibration uncertainties between Swift/XRT and NuSTAR for bright sources, we exclude the Swift/XRT data above 4\,keV and the NuSTAR data below 4\,keV during the fitting process\footnote{Please see more details in \url{http://iachec.scripts.mit.edu/meetings/2019/\\presentations/WGI\_Madsen.pdf} and \url{http://iachec.org/wp-content/presentations/2020/Xcal\_swift\_nustar.pdf}}. The energy bands used for Swift and NuSTAR are 0.5--4\,keV and 4--79\,keV, respectively. We do not perform a joint fit of Insight-HXMT with NuSTAR/Swift. This is because Insight-HXMT, unlike NuSTAR and Swift, is not a focusing telescope, and different background estimation methods may introduce discrepancies in cross-calibration. Additionally, considering that this source is particularly bright and Crab is not visible for Insight-HXMT in June, there are some uncertainties in the calibration of Insight-HXMT during that month. Therefore, we exclude the first two data sets from Insight-HXMT in the fitting. The adopted energy bands for Insight-HXMT are 2--10\,keV for LE and 10--35\,keV for ME. Due to the source's softness and the low signal-to-noise ratio of the HE (35--150\,keV) data, the HE data are not utilized. In the subsequent sections, the results from NuSTAR/Swift and Insight-HXMT will be separately presented. All spectra are fitted within {\tt XSPEC v12.12.0}. Abundances are set to WILM \cite{2000Wilms}, and cross-Asections to VERN \citep[]{1996Verner}.

\subsection{Spectral fitting}
\label{sec:3.1}

Considering that the main emission component of this source is in the soft energy range, the spectral fitting initially uses an absorbed multi-color blackbody model ({\tt model 1: Tbabs*diskbb}) \citep{1986Makishima}. Additionally, a multiplicative constant model ({\tt constant}) is included to account for the normalization discrepancy between different telescopes. However, this model reveals a significant positive residual in the hard X-ray range. Consequently, a thermal Comptonization component ({\tt nthcomp}) \citep{1996Zdziarski} is introduced to describe the hard continuum. The seed photon temperature ($kT_{\rm bb}$) in {\tt nthcomp} is linked to the inner disk temperature ($kT_{\rm in}$) of {\tt diskbb}. Figure~\ref{fig:ratio} shows a soft excess below 1\,keV in the Swift spectra, as well as excesses in the 6--7\,keV and approximately 10--20\,keV bands of the NuSTAR spectra, even though the {\tt diskbb} parameters of the two telescopes are allowed to be different in consideration of the calibration differences between Swift and NuSTAR.

The presence of positive residuals in the 6--7\,keV and 10--20\,keV ranges resembles reflection features caused by disk illumination, either from the corona radiation or the returning photons from the disk. Typically, the former scenario occurs in the hard or intermediate state of BHBs, while the latter is commonly observed in the soft state \citep[e.g.,][]{2020Connors,2021Connors,2021Lazar,2021Wang}. Since 4U 1543--47 is currently in the soft state, we include the reflection model with a blackbody as the incident spectrum, specifically {\tt relxillNS} \citep{Garcia2022}, in the fitting process ({\tt model 3: Tbabs*(diskbb+Nthcomp+relxillNS)}). The {\tt relxillNS} model serves as an intermediary representation of the reflection spectrum formed when the returning disk radiation illuminates the accretion disk. Its primary purpose is to describe the reflection phenomenon originating from accretion disks around neutron stars. Unlike a multi-temperature disk blackbody spectrum, the {\tt relxillNS} model adopts a single-temperature blackbody irradiating spectrum. As a result, its usage demands careful attention to its constraints. Nevertheless, it offers an approximation that proves effective, as the lightbending effects have noticeable impact solely in the extremely inner regions of the accretion disk. In addition, we also attempted the reflection model from the Compton component ({\tt Tbabs*(diskbb+Nthcomp+relxillCp*nthratio})). {\tt nthratio}\footnote{https://github.com/garciafederico/nthratio} is a multiplicative {\sc XSPEC} model that can correct to first-order the unphysical soft excess introduced by {\tt relxillcp} because of its fixed seed photon temperature of 0.01\,keV. However, there are still some noticeable residuals below 1.5\,keV and in the 6--7\,keV and 10--20\,keV energy bands, as depicted in Figure~\ref{fig:relxillcp}. Therefore, we ultimately choose {\tt model 3} to fit the data used in this study. In {\tt model 3}, $kT_{\rm in}$ of {\tt diskbb} and the incident blackbody temperature ($kT_{\rm bb}$) of {\tt relxillNS} are always linked. However, we also conducted initial fits with $T_{\rm in}$ untied from $kT_{\rm bb}$ and found that the other parameters were similar to the linking case. The emissivity indices ($q_{\rm in}$ and $q_{\rm out}$) of {\tt relxillNS} are tied together, and the reflection fraction ($R_{\rm f}$) is fixed at $-1$, indicating that only the blackbody reflected component is considered. The BH spin ($a_*$) is fixed at 0.67 \citep{2020Dong}. It is worth noticing that there is a discrepancy in the spin measurements \citep{2006Shafee,2009Miller,2014Morningstar,2020Dong}, we thus also perform  initial fits by allowing $a_*$ to vary freely and fixing $R_{\rm in}$ at the ISCO radius ($R_{\rm ISCO}$). The resulting spin value is $0.63_{-0.09}^{+0.04}$, which is consistent with that used in this paper ($a_* \sim 0.67$) \citep{2020Dong} and the latest result ($a_* \sim 0.65$) suggested by \citet{2023Yorgancioglu}. Additionally, the other parameters remain essentially unchanged. The iron abundance ($A_{\rm{Fe}}$) and inclination angles ($i$) of these five data sets are linked, respectively. Similarly, the absorption column density ($N_{\rm H}$) is also linked across the data sets, while the remaining parameters are free.

The best-fitting parameters are listed in Table~\ref{tab:sw_nu_dbb_ilink} and the persistent spectra are shown in Figure~\ref{fig:NS_wtga}. The reduced $\chi^2$ values are $\sim$1.2, indicating that additional factors not considered in the reflection model may contribute to the broadening of the iron line, such as a re-emission from disk winds \citep{2004Kallman}, the Compton scattering in the disk atmosphere \citep{2017Steiner}, or an unblurred reflection emission. The residuals at $\sim$2.3\,keV may arise from systematic uncertainties in the Si and Au edges\footnote{\url{https://www.swift.ac.uk/analysis/xrt/digest\_cal.php\#res}}.

Following the spectral fitting of Swift and NuSTAR, we use {\tt model 3} to fit the last three data sets from Insight-HXMT. We fix $N_{\rm H}$ at $\rm{4.0~\times~10^{21}~cm^{-2}}$ for these data sets due to the absence of a low energy data (<2\,keV). $R_{\rm in}$ is fixed at $R_{\rm ISCO}$. Data Set 3 and Data Set 4 contain very few high-energy photons, making it difficult to constrain the normalization of the {\tt diskbb} component. We notice that the normalization values of {\tt diskbb} for the last three data sets in Table~\ref{tab:sw_nu_dbb_ilink} are quite similar. Therefore, we decide to fix the normalization of {\tt diskbb} in Data Set 3 and Data Set 4 at 4000. Moreover, the values of $\Gamma$ for these three data sets are fixed at 3.4, 3.4, and 2, respectively, based on quasi-simultaneous observations as shown in Table~\ref{tab:sw_nu_dbb_ilink}. The fitting results for the Insight-HXMT spectra are shown in Figure~\ref{fig:hxmt_cp_ns}, indicating a good fit. The corresponding best-fitting parameters are presented in Table~\ref{tab:hxmt_dbb_ilink}. The small residuals around 8--10\,keV for Data Set 3 and 4 in Figure~\ref{fig:hxmt_cp_ns} may be caused by background since the LE spectra of Insight-HXMT above 8\,keV are dominated by the background.
 
Since 4U 1543--47 went through a very bright outburst, it is unlikely that its accretion disk remained geometrically thin \citep{2023Yorgancioglu}. Therefore, we also try fitting the Swift/XRT and NuSTAR spectra with the slim disk model {\tt slimbh} \citep{2011Sadowski}, designed for high accretion rates. We fix the BH mass to 9.4 $M_\odot$ , the BH spin parameter to 0.67, the $\alpha$ viscosity parameter to 0.01 \citep{2023Yorgancioglu}, and the distance to 7.5\,kpc. We select a spectral hardening parameter $f_{\rm h}$ to $-1$, which means that the value of $f_{\rm h}$ comes from the TLUSTY\footnote{http://tlusty.oca.eu/index.html} spectra. The limb darkening switch is disabled ($lflag$ = --1), while the raytracing switch is enabled ($rflag$ = 1), allowing for raytracing calculations to be performed from the photosphere while considering the vertical thickness of the disk. Additionally, we tie the inclination of {\tt slimbh} to the inclination of {\tt relxillNS}.

\begin{figure}
    \centering
    \includegraphics[scale=0.5]{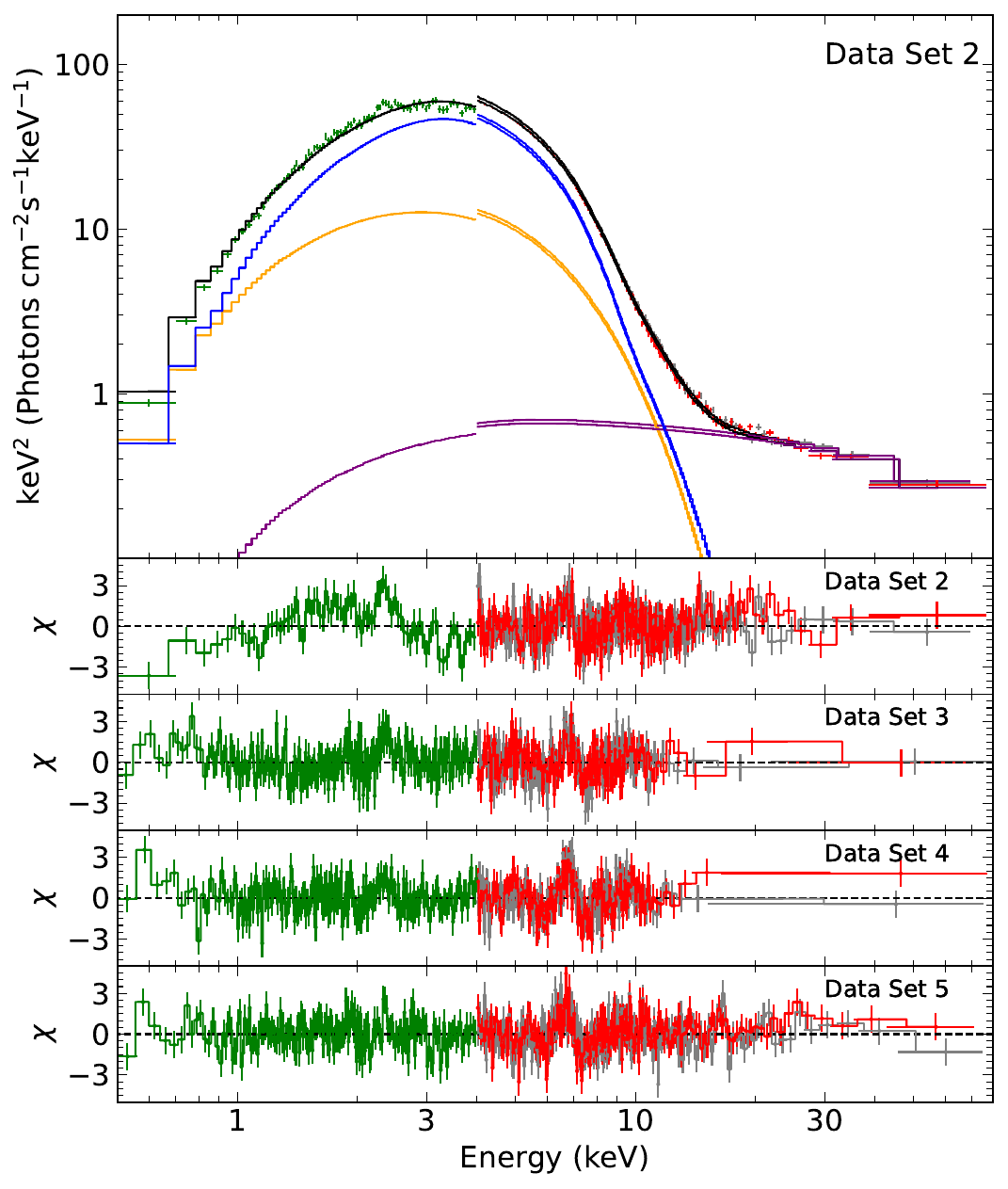} \\
    \caption{Spectra, model components, and spectral residuals for Swift/XRT (green) and NuSTAR (red and gray) with {\tt model 5} ({\tt tbabs*(slimbh+nthcomp+relxillNS)}). The black solid line is the total model, and the orange, purple, and blue solid lines show the {\tt slimbh}, {\tt nthcomp}, and {\tt relxillNS} components, respectively.\\}
    \label{fig:slimbh}
\end{figure}

The remaining parameters, namely the total disk luminosity $L_{\rm disk}$ (in Eddington units) and the normalization parameter, determine the observed spectral shape and flux of the disk emission. The value of the normalization parameter is the fraction of emitted disk photons that reach infinity rather than self-irradiate the disk. The user notes for {\tt slimbh} recommended fixing the normalization to 1. However, we find that this choice always leads to smaller intrinsic values of $L_{\rm disk}$ compared with the observed luminosity. This is because the model normalization of 1 ignores that a fraction of the intrinsic $L_{\rm disk}$ emission is lost to self-irradiation. Instead, we choose to fix $L_{\rm disk}$ and leave the normalization as a free parameter, under the assumption that the proportion of disk photons returning to the disk remains constant at all energy levels. We use the total luminosity $L_{\rm X}$ obtained from each data set (Table~\ref{tab:sw_nu_dbb_ilink}) as input for the corresponding $L_{\rm disk}$ parameter within the {\tt slimbh} model. We neglect photon hardening since the relative contribution of hard photons is minimal. We fix the inclination angle to 30$^{\circ}$, in agreement with the values found from the {\tt diskbb} model fits (Table~\ref{tab:sw_nu_dbb_ilink}). Then, we estimate the observed and unabsorbed luminosities with the convolution model {\tt cflux}. We notice that the {\tt cflux}-derived values of the unabsorbed luminosity are $\approx$1.2 times the input value of $L_{\rm disk}$ set in the simulation. Therefore, we divide the fitted values of $L_{\rm X}$ by this correction factor. Finally, we freeze the $L_{\rm disk}$ parameter for the five data sets at values of 1.4, 0.8, 0.4, 0.33, and 0.30, respectively. The intrinsic luminosity in Data Set 1 exceeds the upper parameter limit of {\tt slimbh} (Eddington limit), hence we exclude this data set from our {\tt slimbh} analysis. Figure~\ref{fig:slimbh} and Table~\ref{tab:sw_nu_slimbh_ilink} show the detailed fitting results.

%%%%%%%%%%%%%%%%%%%%%%%%%
%%%%%%%%%%%%%%%%%%%%%%%%%%%%%

\subsection{Spectral results}
By fitting the spectra of NuSTAR/Swift and Insight-HXMT, we find that the complicated spectra of the 2021 outburst can be well described by the disk reflections from the returning disk photons and a weak corona. As shown in Table~\ref{tab:sw_nu_dbb_ilink} and Table~\ref{tab:hxmt_dbb_ilink}, the disk temperature $T_{\rm in}$ gradually decreases from around $\sim 1.1$\,keV near the outburst peak (Data Set 1) to approximately $(0.8-0.9)$\,keV at the end of the soft state (Data Set 5). This trend is in agreement with the gradual decrease in the accretion rate since the outburst peak. The photon index $\Gamma$ of the Comptonization component is $\gtrsim$ 2, as expected for a soft source. The iron abundances $A_{\rm Fe}$ are also found to be relatively consistent across both tables. However, we also observe that some parameters (e.g., log~$\xi$ and log~$\rho$) show discrepancies between the fitting values from different satellites. This is mainly attributed to differences in the cross-calibration between the satellites, as mentioned earlier. Therefore, we do not discuss the specific values of these physical parameters further.

The reflection component accounts for most of the total flux (Tables~\ref{tab:sw_nu_dbb_ilink} and \ref{tab:hxmt_dbb_ilink}), which indicates that a large fraction of soft photons returns to the disk. This high reflection fraction is also obtained when we substitute {\tt slimbh} for {\tt diskbb} (Table~\ref{tab:sw_nu_slimbh_ilink}, see also Section~\ref{sec:4.1}). Moreover, Data Set 1 and 2 (taken at higher luminosity) have lower values of $q_{\rm in}$ ($= q_{\rm out}$) than Data Set 3, 4 and 5.

%__________________________________________________________________

\section{Discussion and conclusion}
\label{sec:dis}

The 2021 outburst of 4U 1543--47 is the brightest among its five known outbursts, reaching up to  $\sim8$ Crab \citep{2021ATel14708}. In this paper, we have plotted the evolution of the hardness ratio using data from MAXI/GSC and HXMT/LE during the outburst, and we find that 4U 1543--47 rapidly transitioned into a soft state at the beginning of the outburst, which is similar to its outbursts in 1983 \citep{1984Kitamoto} and 2002 \citep{2004Park,2020Russell}. Subsequently, we focus on analyzing the spectra of five data sets obtained from quasi-simultaneous observations using Insight-HXMT, NuSTAR, and Swift during the 2021 outburst. Remarkably, despite the source being in a soft state, we find a prominent reflection component (more than half of the total flux), a phenomenon rarely reported for black hole systems. Based on these observations, our next discussion will revolve around the accretion geometry of 4U 1543--47. 

\subsection{Geometrically thick disk}
\label{sec:4.1}

\begin{figure}
    \centering
    \includegraphics[scale=0.32]{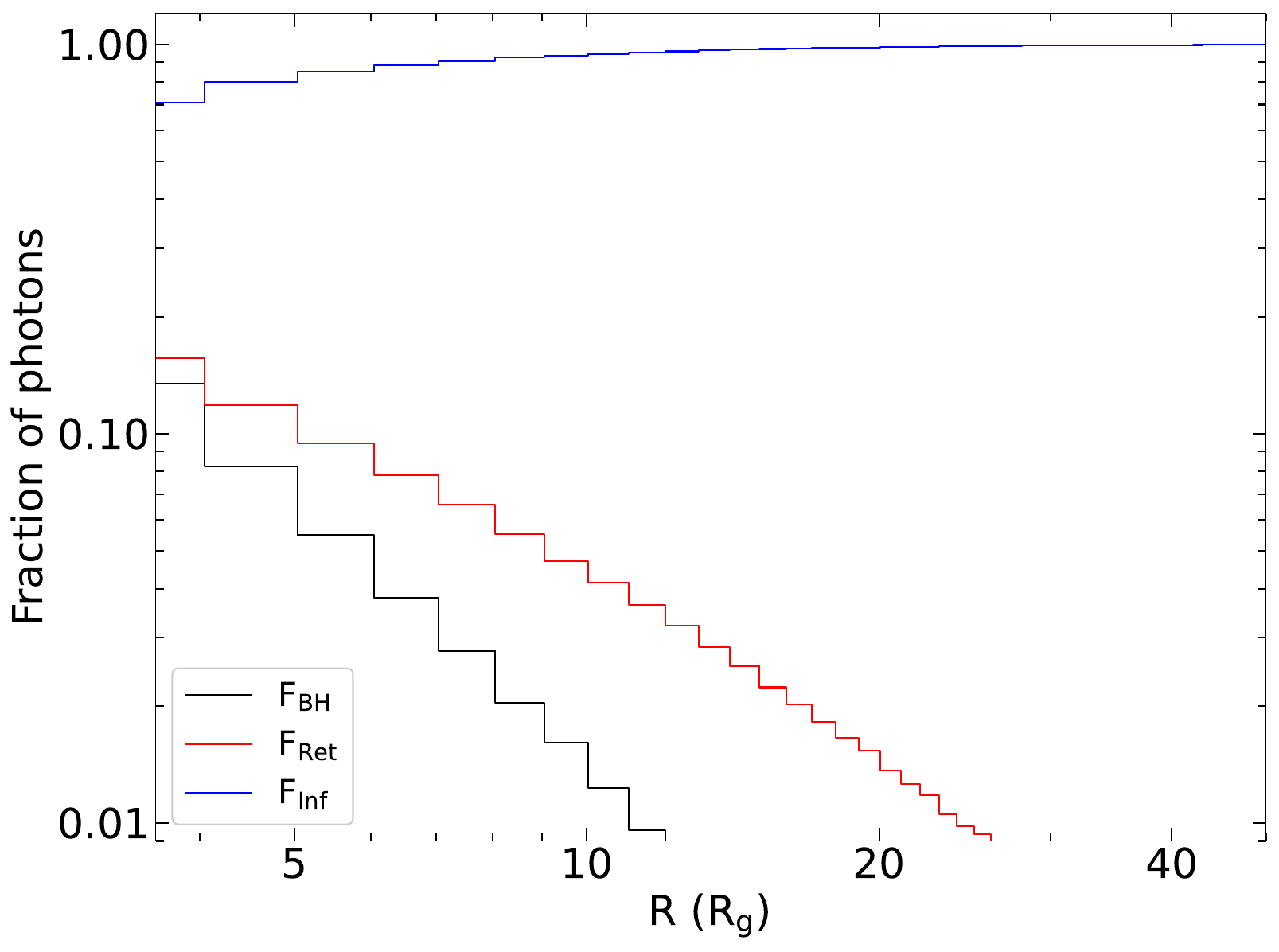}
    \caption{Radial profiles of fraction of disk photons that fall into the black hole, return to the disk, or reach infinity for a razor-thin disk. The spin of the black hole is set to 0.67. }
    \label{fig:fraction_photons}
\end{figure}

\begin{figure}
    \centering
    \includegraphics[scale=0.32]{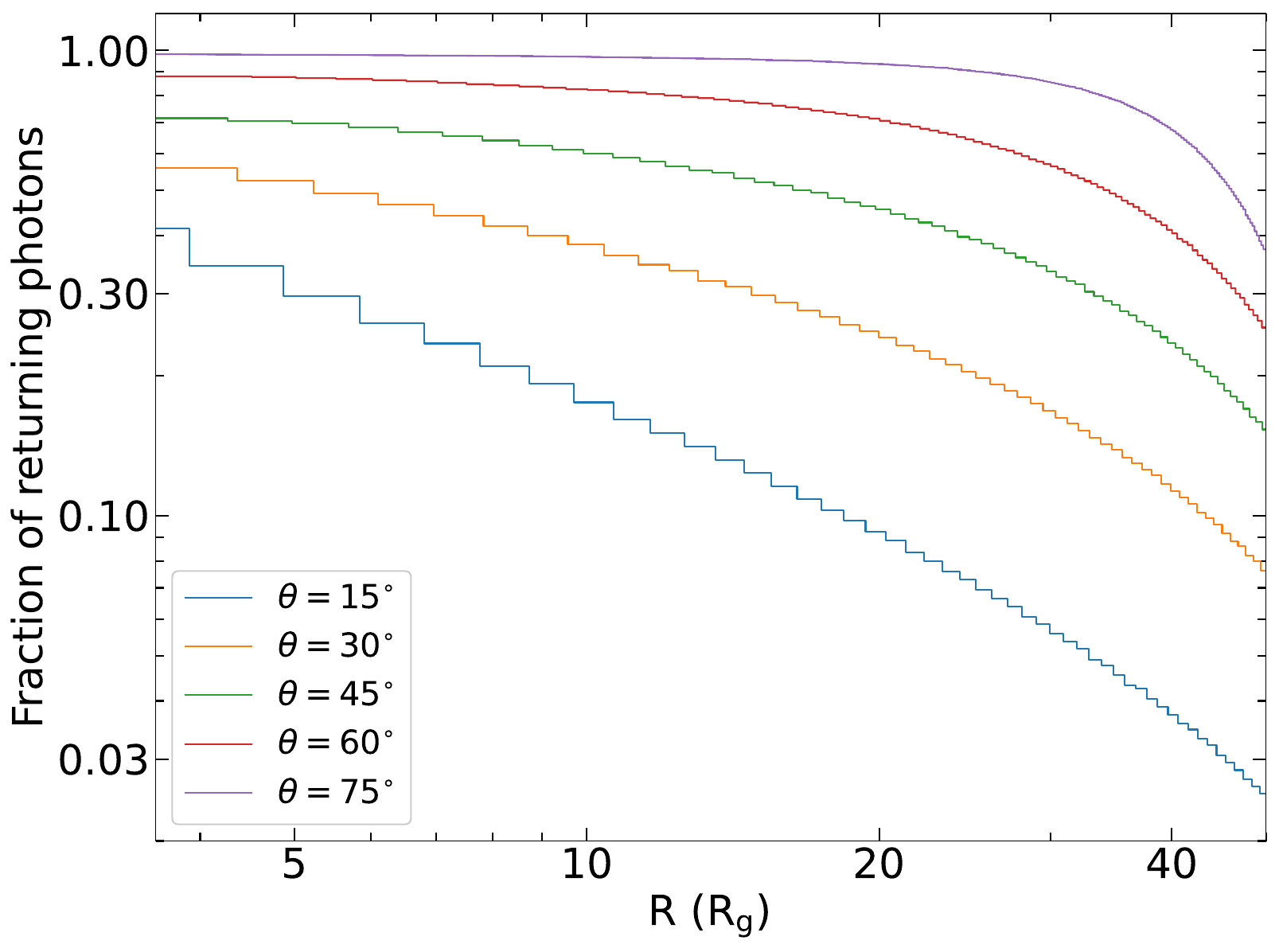}
    \caption{Radial profiles of the fraction of disk photons returning to the disk with different $\theta$ for a funnel-shaped disk. The spin of the black hole is set to 0.67.}
    \label{fig:returning_fraction}
\end{figure}
Although the theory of disk self-irradiation has been proposed for a long time \citep{1976Cunningham}, it was only in 2020 that \cite{2020Connors} presented the first evidence for the existence of returning radiation from the thermal disk radiation, producing the relativistic reflection in the soft state of the black hole XTE J1550--564. Subsequently, in 4U 1630--47 \citep{2021Connors} and EXO 1846--031 \citep{2021Wang}, the reflection was also reported during the soft state and effectively modeled using {\tt relxillNS}. However, the fractional flux (relative to the total flux) of the reflected component in these sources is not high, consistent with the estimated fraction of photons returning to the disk under the assumption of a \cite{1973Shakura} disk (thin disk). Recently, a high fraction of returning radiation was found in the soft state spectrum of MAXI J1631--479 \citep{2023Rout}, with the flux of $\sim1$ Crab in the 2--10\,keV range \citep{2021Rout}. However, a detailed investigation of this phenomenon has not been conducted yet. To verify whether the high reflection fraction observed in the spectra of 4U 1543--47 could be attributed to disk self-irradiation, we conducted general relativity ray-tracing simulations using SIM5 \citep{2017Bursa,2013Pontzen}. In our simulations, we set the black hole spin and the outer radius of the disk to 0.67 and 50\,$R_{g}$ ($R_{\rm g}=GM/c^{2}$, where $M$ is the BH mass, $G$ is the gravitational constant and $c$ denotes the speed of
light) \citep{2011Dotan,2014Jiang}, respectively. Based on the fitting results of {\tt relxillNS} in Section~\ref{sec:res}, we set the inner radius of the disk at the ISCO. The photons are emitted according to the disk geometry (in the frame of rest of the disk), where the number of photons emitted at each emission angle is proportional to $\cos{\Theta_{\rm e}}$, where $\Theta_{\rm e}$ is the polar emission angle defined with respect to the normal of the disc surface. We initially assumed a razor-thin disk with the coordinate angular velocity as a function of radius $R$ in units of $R_{\rm g}$, satisfying $\omega_{0}=1/(R^{3/2}+a_*)$. We then calculate the fraction of photons that fall into the black hole, return to the disk, or escape to infinity (Figure~\ref{fig:fraction_photons}). The number of photons emitted per unit area per unit time by a blackbody with temperature $T$ is proportional to $T^3$. Assuming a disk temperature $T \propto R^{-3/4}$, we sum up the photons returning to the disk at different radii. However, the total fraction was approximately $\sim$6.0\%, much smaller than the result observed ($\gtrsim 50\%$) in our study.

Considering the high accretion rate of 4U 1543--47, the disk is likely to be geometrically thick. Therefore, we perform ray-tracing simulations assuming a "funnel-like" thick disk, where the angle between the disk surface and the equatorial plane is $\theta$. When $\theta = 0^{\circ}$, it corresponds to the razor-thin disk described above. Moreover, we assume the coordinate angular velocity of the disk surface is $\omega_{\theta}$ = $\omega_{0}~\times~\cos^{3/2}{\theta}$. We obtain radial profiles of the fraction of disk photons returning to the disk corresponding to different values of $\theta$ as shown in Figure~\ref{fig:returning_fraction}. Based on the above discussion, we find that the total fraction of photons returning to the disk is approximately 20.4\%, 38.4\%, 57.3\%, 76.3\%, and 91.4\% for $\theta$ values of 15, 30, 45, 60, and 75 degrees, respectively, when assuming $T \propto R^{-3/4}$. In a possible thick disk scenario where $T \propto R^{-1/2}$, the total fraction of photons is also approximately the same for the same $\alpha$ values. Thus, when $\gtrsim45^{\circ}$, in the case of a geometrically thick disk, it is expected that the returning radiation will constitute $\gtrsim$60\% of the total flux. This can be used to explain the observational results of 4U 1543--47. 
{\tt model 4} provides a good fit to the spectra (Table 4 and Figure 7), further confirming the significant self-irradiation of the disk in the case of a geometrically thick disk. Only about 30--40\% of the disk photons are directly emitted to infinity. As shown in Tables~\ref{tab:sw_nu_dbb_ilink} and \ref{tab:sw_nu_slimbh_ilink}, with increasing luminosity, the emissivity index $q_{\rm in}$ ($=q_{\rm out}$) becomes flatter, which is consistent with a scenario involving a higher accretion rate and thicker accretion disk with a larger angle between the disk surface and the equatorial plane.

Furthermore, this "funnel-like" disk geometry is also consistent with the GRRMHD simulation results in both super-critical and sub-critical states \citep[e.g.][]{2014Sadowski,2014Jiang,2022Wielgus,2023Huang}. Another conclusion drawn from the simulations is that a naturally formed corona with high temperature and low density appears above the black hole in high accretion rate scenarios \citep{2023Huang}. This corona interacts with soft photons through inverse Compton scattering, resulting in a PL component. The high-energy PL tail observed in the spectra of 4U 1543--47 may serve as evidence for the existence of such a corona. In addition, considering the significant fraction ($\gtrsim$ 50\%) of photons returning to the disk, we expect a substantial heating effect on the disk surface, potentially contributing to the formation of radiatively driven outflows.

\subsection{Other explanations for the reflection-like residuals in the soft state}
\label{sec:4.2}

Recent investigations on this outburst, such as \citet{2023Draghis}, \citet{2023Prabhakar} and \cite{2023Husain} have proposed different physical interpretations. All of these studies used the reflection from the Comptonization component (i.e., {\tt relxill} or {\tt relxilllp}) to fit the spectra and found that either a model for partial covering of the source with ionized absorbing material \citep[{\tt zxipcf;}][]{2023Draghis,2023Husain} or a Gaussian absorption model \citep[{\tt gabs;}][]{2023Prabhakar} is required to improve the fit and reduce the spectral residuals. This absorption feature is believed to be associated with the presence of disk winds. However, based on our fitting results and the findings in \cite{2023Prabhakar}, we observe that the corona only contributes to a small fraction (less than 8\%) of the total flux, suggesting that the illumination from such a weak corona should be correspondingly faint.

We also note that the radiation from the plunging region located inside the ISCO is proposed to explain the reflection-like features of MAXI J1820+070 and MAXI J0637--430 in the soft state \citep{2020Fabian,2021Lazar}. However, the plunging region was considered to have a negligible contribution to the energy spectrum due to the small plunging time scale \citep[][]{Page1974}. Although the magnetohydrodynamic simulation results show that the plunging region could generate a blackbody emission in the non-zero torque case \citep{2012Zhu}, this effect is prominent only when the BH spin is low, since the ISCO is relatively far from the event horizon. In view of this, the possible emission of the plunging region may be the cause of the spectral features of MAXI J1820+070 and MAXI J0637--430, as their spin is $a_{*}\sim$0.2 \citep{Guan2021} and $a_{*} \lesssim$0.25 \citep{Soria2022}, respectively. But for 4U 1543--47, by considering the revised spin value from the continuum fitting \citep{2023Yorgancioglu} and the spin from the reflection modeling \citep{2020Dong}, it has a moderately high spin of $\gtrsim$0.7, we thus expect that the emission from the plunging region might be insignificant.

%__________________________________________________________________
\begin{acknowledgements}
We thank Riley M. T. Connors for useful comments that have improved this paper. This work made use of data from the Insight-HXMT mission, a project funded by China National Space Administration (CNSA) and the Chinese Academy of Sciences (CAS), and the Swift data supplied by the UK Swift Science Data Centre at the University of Leicester. This work is supported by the National Key R\&D Program of China (2021YFA0718500). We acknowledge funding support from the National Natural Science Foundation of China (NSFC) under grant Nos. 12122306 and 12073029, the CAS Pioneer Hundred Talent Program Y8291130K2 and the Scientific and technological innovation project of IHEP Y7515570U1. YZ acknowledges support from the China Scholarship Council (CSC), no.\ 201906100030, and the Dutch Research Council (NWO) Rubicon Fellowship, file no.\ 019.231EN.021.
\end{acknowledgements}

%-------------------------------------------------------------------

\bibliographystyle{aa}
\bibliography{ref}

%-------------------------------------------------------------------
\begin{appendix}
\section{The probability distributions of the parameters from Data Set 3}
\label{sec:app}

\begin{figure*}
    \centering
    \includegraphics[width=1\textwidth]{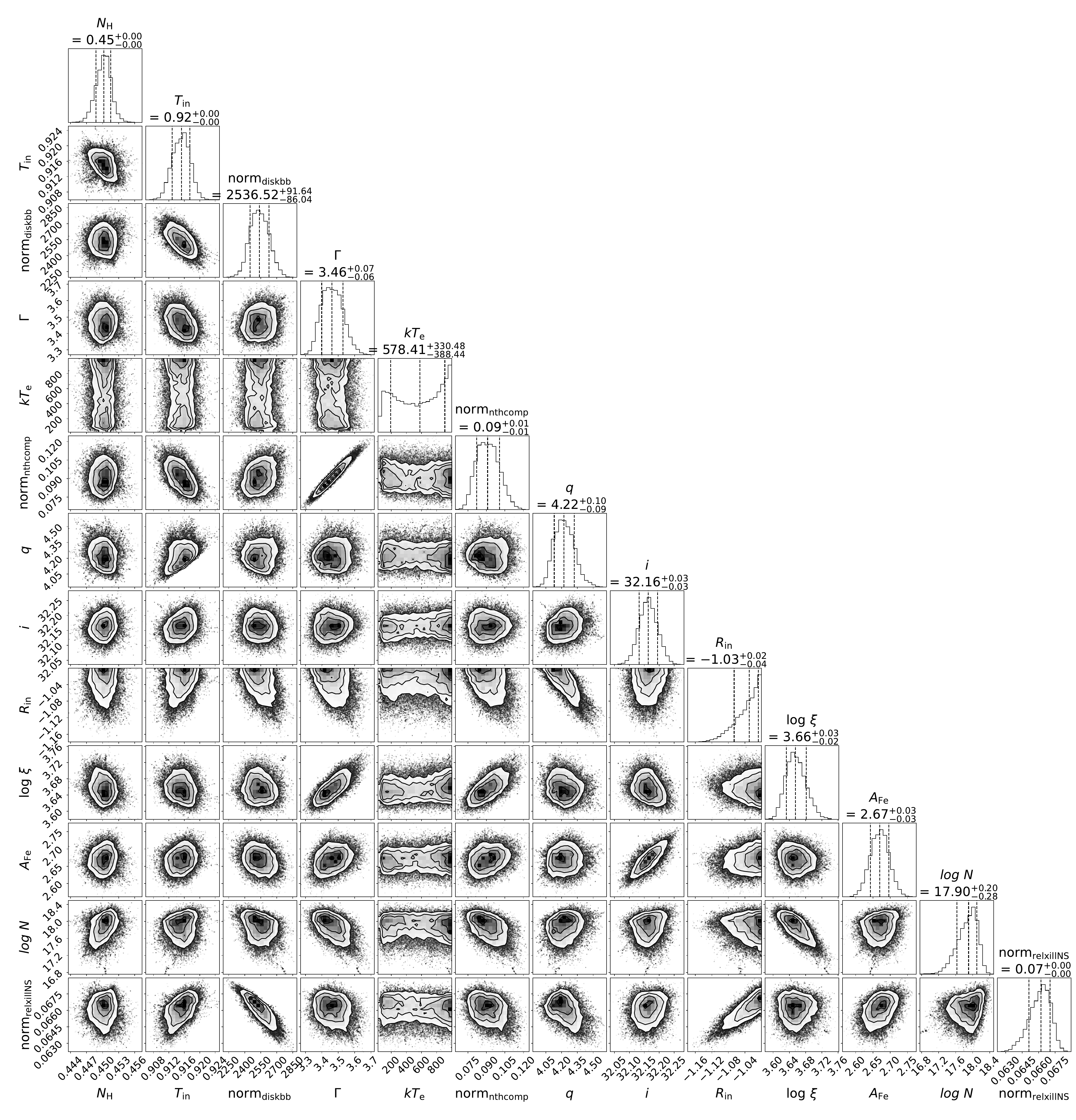}
    \caption{The probability distributions of the parameters for {\tt model 3} obtained from the Swift/XRT and NuSTAR Data Set 3 through Markov chain Monte-Carlo algorithm (MCMC). }
    \label{fig:corner1}
\end{figure*}

\begin{figure*}
    \centering
    \includegraphics[width=1\textwidth]{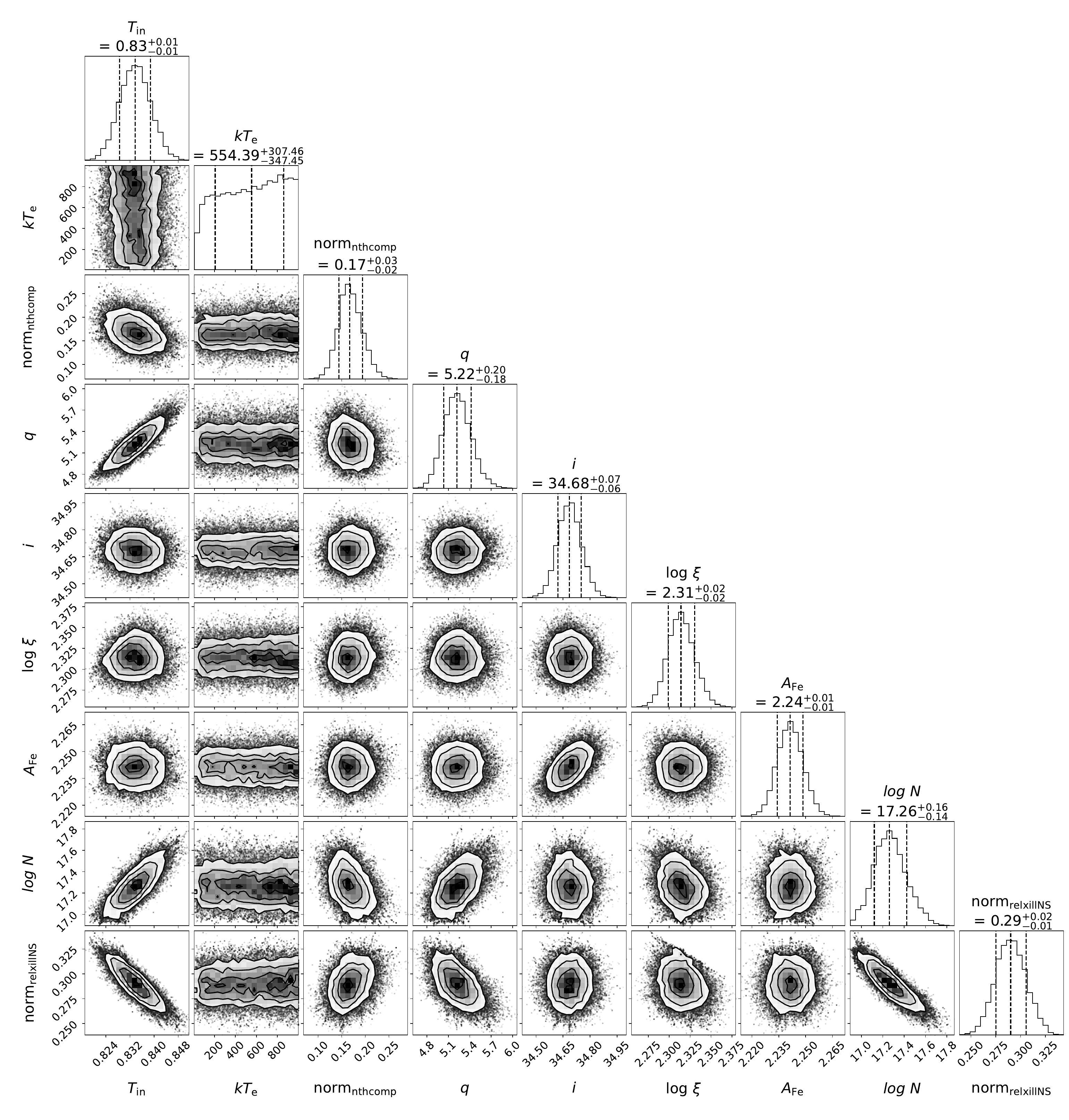}
    \caption{The probability distributions of the parameters for {\tt model 3} obtained from the Insight-HXMT Data Set 3 through MCMC. }
    \label{fig:corner2}
\end{figure*}

\begin{figure*}
    \centering
    \includegraphics[width=1\textwidth]{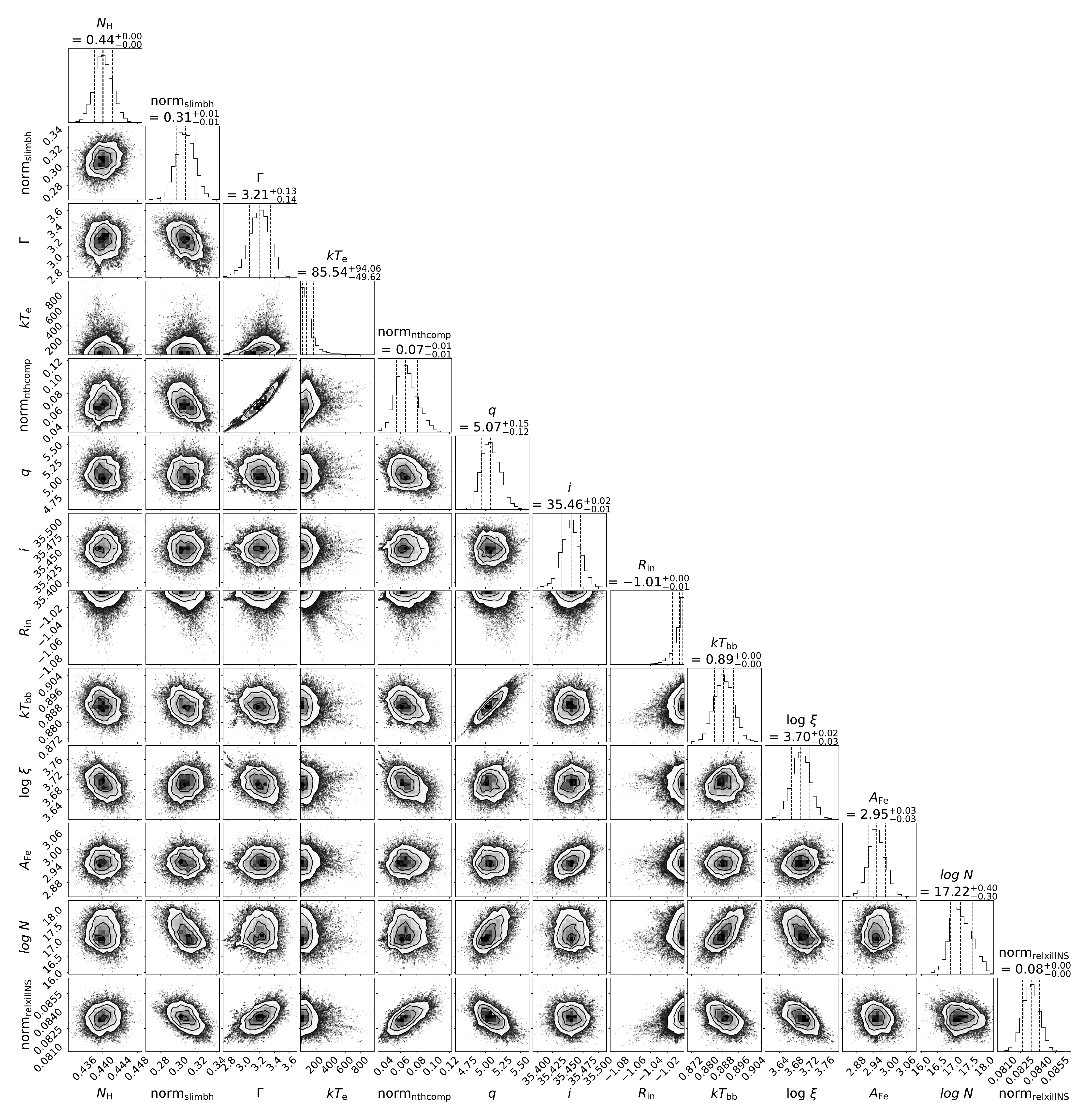}
    \caption{The probability distributions of the parameters for {\tt model 4} obtained from the Swift/XRT and NuSTAR Data Set 3 through MCMC. }
    \label{fig:corner3}
\end{figure*}

\end{appendix}

\end{document}